\documentclass[twocolumn]{aastex631}
\usepackage{amsmath}
\usepackage{float}
\usepackage{graphicx}
\usepackage{booktabs}

\begin{document}

\title{The $\delta$ Scuti pulsator occurrence as a function of age, $T_{\rm eff}$, rotation, and metallicity}

\correspondingauthor{Ian Berry}
\email{ianberry@hawaii.edu}

\author[0009-0009-3641-4137]{Ian Berry}
\affiliation{Institute for Astronomy, University of Hawai`i, 2680 Woodlawn Drive, Honolulu, HI 96822, USA}

\author[0000-0001-8832-4488]{Daniel Huber}
\affiliation{Institute for Astronomy, University of Hawai`i, 2680 Woodlawn Drive, Honolulu, HI 96822, USA}

\author[0000-0003-3020-4437]{Yaguang Li}
\affiliation{Institute for Astronomy, University of Hawai`i, 2680 Woodlawn Drive, Honolulu, HI 96822, USA}

\author[0000-0003-3244-5357]{Daniel Hey}
\affiliation{Institute for Astronomy, University of Hawai`i, 2680 Woodlawn Drive, Honolulu, HI 96822, USA}

\author[0000-0001-5222-4661]{Timothy R. Bedding}
\affiliation{Sydney Institute for Astronomy, School of Physics, University of Sydney, NSW 2006, Australia}

\author[0000-0002-5648-3107]{Simon J. Murphy}
\affiliation{Centre for Astrophysics, University of Southern Queensland, Toowoomba, QLD 4350, Australia
}

\begin{abstract}
    Many A- and F- type stars do not display $\delta$ Scuti pulsations, despite being located within the instability strip. We use photometry from the TESS Mission to discover and study $\delta$ Scuti pulsators in open clusters within 500 pc and with ages between ~20 and 900 Myr, which provide a unique opportunity to study $\delta$ Scuti pulsators in coeval populations with uniform chemical composition. We measure pulsator occurrence, which corrects the pulsator fraction for incompleteness, across all clusters. We find that clusters younger than 200 Myr tend to exhibit higher occurrence rates, with an average occurrence of 88$\pm$3\%. The occurrence rates in clusters older than 200 Myr tend to resemble the pulsator fraction of field-star samples, with an average occurrence of 62$\pm$3\%. In addition, we find that pulsators tend to rotate more rapidly in older clusters than their younger counterparts and that hotter pulsators may stop pulsating earlier than their cooler counterparts. These results show that pulsator occurrence decreases with age and that rapid rotation is critical in maintaining $\delta$ Scuti pulsations over time.

\end{abstract}


\section{Introduction}

The $\delta$ Scuti pulsators are spectral type A and F stars on or near the main sequence within the classical Cepheid instability strip, spanning effective temperatures ($T_{\rm eff}$) between $\sim7000$ and $9500$ K. These stars display high-frequency pressure (p) modes due to internal helium ionization zones, which drive radial and non-radial pulsations through periodic changes in opacity known as the $\kappa$ mechanism \citep{2004A&A...414L..17D}. Some $\delta$ Scuti pulsators also show low-frequency gravity (g) modes associated with the $\gamma$ Doradus pulsators, and are known as hybrid pulsators \citep{2010AN....331..989G, 2010arXiv1007.3176H,2014MNRAS.444..102K,2018FrASS...5...43B}.


\begin{figure*}
    \centering
    \includegraphics[width=.80\linewidth]{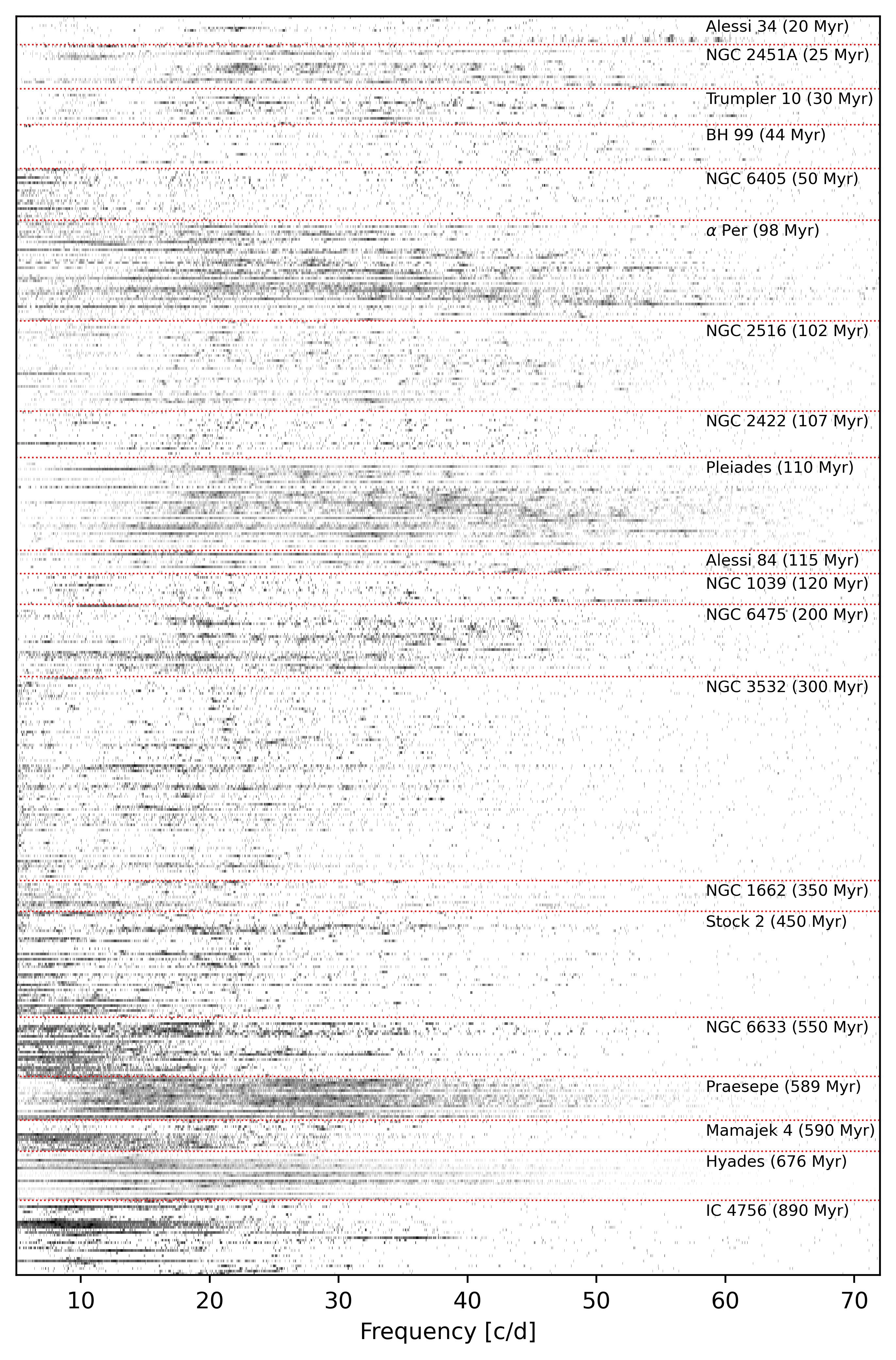}
    \caption{Stacked amplitude spectrum of all $\delta$ Scuti pulsators in this study. Clusters are ordered by age (youngest on top). The red dotted lines separate each cluster. Within each cluster, stars are sorted by $T_{\rm eff}$, with the coolest on top. Note the appearance of $\delta$ Scuti stars with regular pulsation patterns in the younger ($\lesssim$ 200 Myr) open clusters.}
    \label{fig:stacked_amp_spec}
\end{figure*}



The physics of the driving mechanism of pulsations in $\delta$ Scuti stars remains poorly understood, with one issue being that some stars within the instability strip seem to lack pulsations entirely, despite theoretical expectations \citep{2004A&A...414L..17D,2005A&A...435..927D}. In addition, stars that appear to fall outside the bounds of the instability strip can show $\delta$ Scuti pulsations \citep[e.g.][]{2011A&A...534A.125U,2018FrASS...5...43B,2019MNRAS.485.2380M,2023A&A...674A..36G}.

As a consequence, much work has been done to measure the pulsator fraction, which is the ratio of pulsators to the total sample size, which appears to vary between different stellar populations. Among field stars, the pulsator fraction varies between 50 and 70\% \citep{2019MNRAS.485.2380M,2024MNRAS.528.2464R,2025MNRAS.542.2866M}. In general, higher fractions have been measured among brighter stars than among dimmer stars \citep{2024MNRAS.528.2464R,2025MNRAS.542.2866M}, suggesting that detection biases/capabilities could affect whether certain instability strip stars appear to pulsate.

The concept of ``pulsator occurrence'' was introduced by \citet{2025ApJ...995..128B}, who used injection and recovery tests and a known $\delta$ Scuti amplitude distribution from \citet{2019MNRAS.485.2380M} to correct  the pulsator fraction for detection biases. This was done using instability strip stars in the 300 Myr old open cluster NGC 3532, where a 63\% pulsator occurrence was measured, up from the pulsator fraction of 50\%.


Physical factors must also determine whether a star within the instability strip pulsates. The driving of $\delta$ Scuti pulsations through the $\kappa$ mechanism requires the presence of a helium ionization zone at a specific depth and temperature \citep[$30,000-40,000\,{\rm K}$;][]{1996ima..book.....C} in the stellar interior. In slowly rotating stars, helium can diffuse out of the ionization zone over time \citep{1973A&A....23..221B}, stopping pulsations from the $\kappa$ mechanism. For a 1.6${\rm M_\odot}$ star with a stellar wind, 50\% of helium will be depleted from the ionization zone by 100 Myr, and up to 80\% by 500 Myr \citep{2005A&A...443..627T,2016A&A...589A.140D}. Therefore, the youngest instability strip stars may still have the necessary helium in the ionization zone to drive pulsations, regardless of other stellar parameters.

\begin{table*} \label{tab:clusters}
\centering
\caption{Properties of the open clusters used in this work. Equatorial coordinates, distances, and distance uncertainties are taken from \citet{HuntReffert23, HuntReffert24}. $N_{\delta\,{\rm Sct}}$ is the number of $\delta$ Scuti stars. The Age Ref. and [Fe/H] Ref. columns show the source of the age and metallicity for each cluster.}
\begin{tabular}{lccccccccc}
\toprule
Cluster & $\alpha$ [$^\circ$] & $\delta$ [$^\circ$] & Distance [pc] & Age [Myr] & [Fe/H] & $N_{\delta\,{\rm Sct}}$ & Occurrence [\%] & Age Ref. & [Fe/H] Ref. \\
\midrule
Alessi 34   & 120.062 & -50.655 & 485$\pm29$ & $20^{+12}_{-9}$  & 0.09$\pm0.12$ & 11 & 86$\pm14$ & 1 & 2 \\
NGC 2451A   & 115.971 & -38.241 & 190$\pm40$ & $25^{+22}_{-12}$  & -0.11$\pm0.12$ & 17 & 93$^{+7}_{-13}$ & 1 & 3 \\
Trumpler 10 & 131.937 & -42.534 & 428$\pm5$ & $30^{+30}_{-20}$  & -0.12$\pm0.06$ & 14 & 77$\pm13$ & 4 & 5 \\
BH 99       & 159.554 & -59.109 & 440$\pm6$ & $44^{+38}_{-19}$  & 0.08$_{-0.09}^{+0.12}$  & 17 & 100$^{+0}_{-14}$ & 1 & 2  \\
NGC 6405    & 265.067 & -32.228 & 451$\pm8$ & $50^{+38}_{-21}$  & 0.07$\pm0.03$  & 20 & 100$^{+0}_{-15}$ & 1 & 6  \\
$\alpha$ Per & 51.332 & 49.119  & 173$\pm26$ & $98\pm2$ & 0.03$\pm0.08$ & 39 & 69$\pm8$ & 7 & 8 \\
NGC 2516    & 119.521 & -60.793 & 406$\pm17$ & $102\pm15$ & -0.08$\pm0.01$ & 35 & 76$\pm12$ & 9 & 10 \\
NGC 2422    & 114.169 & -14.493 & 468$\pm7$ & $107^{+88}_{-45}$  & -0.05$\pm0.02$ & 17 & 95$^{+5}_{-16}$ & 1 & 10  \\
Pleiades    & 56.680  & 24.108  & 134$\pm9$ & 110$^{+32}_{-28}$ & 0.03$\pm0.05$  & 36 & 86$\pm7$ & 11 & 12 \\
Alessi 84   & 110.510 & 55.384  & 198$\pm27$ & $115_{-50}^{+89}$ & 0.11$\pm0.07$ & 9 & 91$^{+9}_{-16}$ & 1 & 8 \\
NGC 1039    & 40.548  & 42.722  & 489$\pm12$ & $120_{-49}^{+108}$ & 0.00$\pm0.13$  & 12 & 89$^{+11}_{-18}$ & 1 & 8  \\
NGC 6475    & 268.476 & -34.828 & 275$\pm5$ & $200\pm50$ & 0.03$\pm0.02$  & 28 & 73$\pm9$ & 13 & 13 \\
NGC 3532    & 166.394 & -58.694 & 471$\pm8$ & $300\pm100$ & -0.07$\pm0.01$ & 79 & 59$\pm6$ & 14 & 15 \\
NGC 1662    & 72.114  & 10.904  & 404$\pm12$ & $350_{-125}^{+236}$ & -0.03$\pm0.13$ & 12 & 65$\pm14$ & 1 & 8 \\
Stock 2     & 33.850  & 59.577  & 370$\pm7$ & $450\pm150$ & -0.07$\pm0.06$ & 41 & 41$\pm8$ & 16 & 16 \\
NGC 6633    & 276.805 & 6.559   & 389$\pm8$ & $550_{-100}^{+50}$ & -0.15$\pm0.08$  & 23 & 61$\pm10$ & 17 & 8 \\
Praesepe    & 130.088 & 19.666  & 183$\pm3$ & $589_{-26}^{+13}$ & 0.21$\pm0.01$  & 17 & 73$\pm11$ & 18 & 19  \\
Mamajek 4   & 276.371 & -50.637 & 444$\pm13$ & 590$_{-12}^{+10}$ & 0.05$\pm0.07$  & 12 & 49$\pm15$ & 20 & 21\\
Hyades      & 66.709  & 16.081  & 47$\pm6$  & $676_{-30}^{+13}$ & 0.15$\pm0.02$  & 19 & 70$\pm11$ & 18 & 22 \\
IC 4756     & 279.627 & 5.437  & 466$\pm10$ & $890\pm70$ & 0.09$\pm0.13$ & 29 & 67$\pm10$ & 23 & 8 \\
\bottomrule
\end{tabular}

\vspace{3pt}
{\small
\textbf{References:}
(1) \citet{HuntReffert23, HuntReffert24};
(2) \citet{2024AJ....167...12C}
(3) \citet{1983AJ.....88.1769P}
(4) \citet{2023A&A...675A.167P};
(5) \citet{2021A&A...649A..54P};
(6) \citet{2016AJ....151...49K};
(7) \citet{2022MNRAS.513..374P};
(8) \citet{2024A&A...692A.212Z};
(9) \citet{2024A&A...686A.142L};
(10) \citet{2018MNRAS.475.1609B};
(11) \citet{2018A&A...616A..10G};
(12) \citet{2009AJ....138.1292S};
(13) \citet{Villanova_2009};
(14) \citet{2011AJ....141..115C};
(15) \citet{2019A&A...622A.110F};
(16) \citet{2021A&A...656A.149A};
(17) \citet{2025A&A...702A.265B};
(18) \citet{2018ApJ...863...67G};
(19) \citet{2020A&A...633A..38D}
(20) \citet{2025AJ....170..288L};
(21) \citet{2021MNRAS.504..356D};
(22) \citet{2016MNRAS.457.3934L};
(23) \citet{2015A&A...580A..66S};
}

\end{table*}


Rotation can also affect the driving of pulsations. Gravitational settling of helium out of the ionization zone can be counteracted by rotational mixing through rapid rotation \citep{2004A&A...425..591H}. \citet{2024ApJ...972..137G} found a clear increase in pulsator fraction with increasing rotation rate using a large sample of $\delta$ Scuti pulsators. Similar results were found  with $\delta$ Scuti stars in the Cep-Her Complex \citep{2024MNRAS.534.3022M}, the Pleiades \citep{2023ApJ...946L..10B} and NGC 3532 \citep{2025ApJ...995..128B}. 


\begin{figure*}
    \centering
    \includegraphics[width=\linewidth]{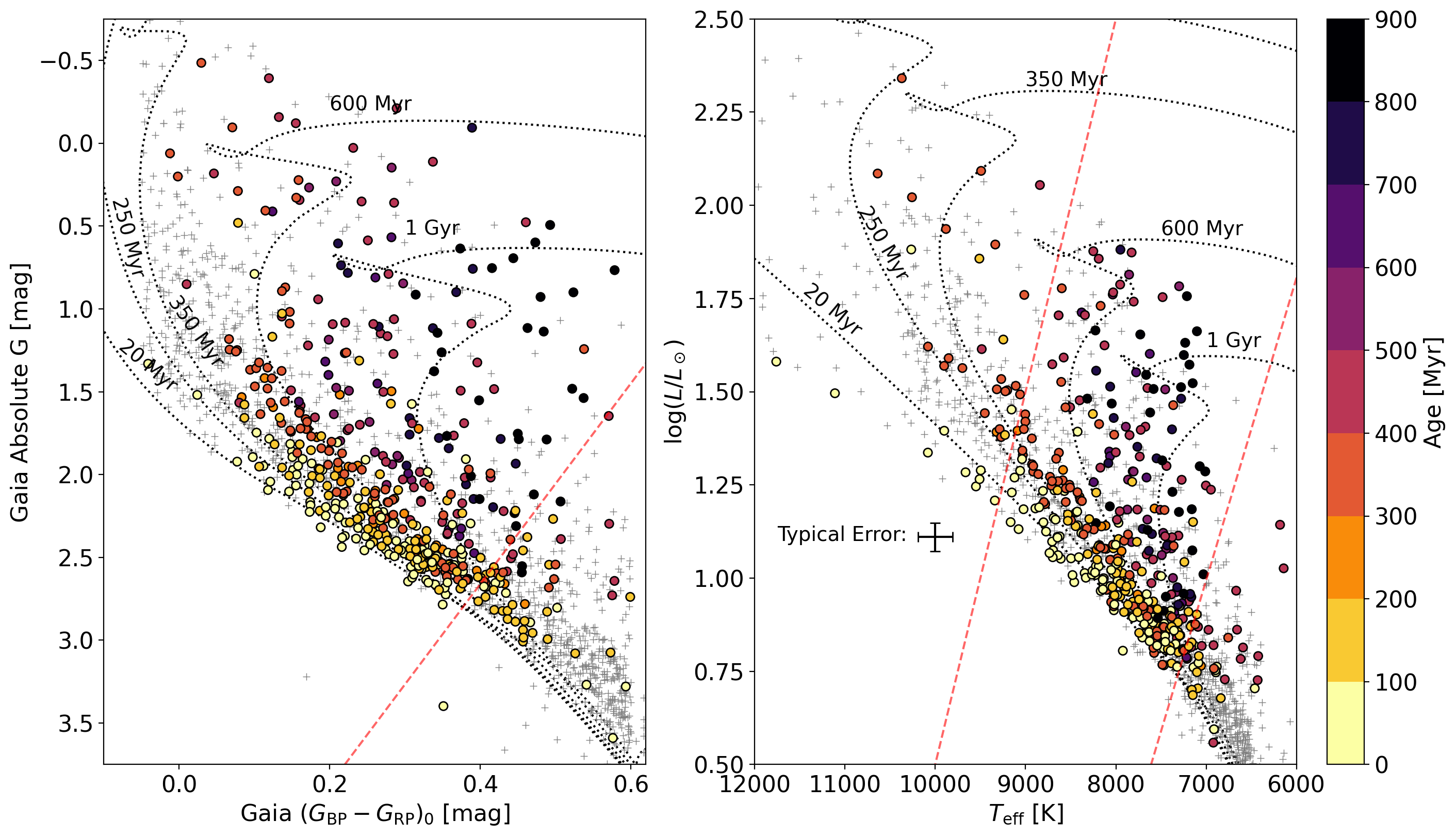}
    \caption{Gaia CMD (left panel) and HR diagram (right panel) for all clusters studied in this work. Color-mapped points are $\delta$ Scuti pulsators, and colors show age. Gray crosses are non-detections. The black dotted curves show 40\% critical rotation MIST isochrones \citep{2016ApJS..222....8D, 2016ApJ...823..102C} at various ages. In the left panel, the red dashed line shows the empirical red instability strip edge from \citet{2024ApJ...972..137G}. In the right panel, the red dashed lines show the empirical bounds of the instability strip from \citet{2019MNRAS.485.2380M}.}
    \label{fig:CMD_HR}
\end{figure*}

Chemical composition can affect the driving of pulsations. The metallic-lined Am stars display strong metal absorption lines, and are generally known to be slow rotators. This is thought to allow helium to gravitationally diffuse out of the ionization zone, and metals to rise to the surface through radiation pressure \citep{1974NInfo..32..104P, 1980LNP...125...22D, 2015A&A...579A.116O}. These stars appear to have lower pulsation fractions than their non-metallic-lined counterparts, about 13\% \citep{1970ApJ...162..597B, 2021arXiv210709479G, 2024A&A...690A.104D}. 


The Transiting Exoplanet Survey Satellite \citep[TESS;][]{2015JATIS...1a4003R} enables a systematic study of $\delta$ Scuti stars in open clusters across nearly the entire sky. With extended-mission full-frame-image (FFI) cadences of 10 minutes and 200 seconds, TESS can detect any $\delta$ Scuti pulsator regardless of pulsation frequency. These FFIs now cover over 90\% of the sky, allowing studies of nearly all nearby ($\lesssim$ 500 pc) $\delta$ Scuti stars in both the field and open clusters.

Open clusters and moving groups are an ideal environment for measuring pulsator fractions at a fixed age and chemical composition. Pulsator fractions and occurrences have been measured with TESS in three young stellar populations and one intermediate age open cluster: the Pleiades \citep[100 Myr;][]{2023ApJ...946L..10B}, NGC 2516 \citep[100 Myr;][]{2024A&A...686A.142L}, the Cep-Her complex \citep[$\leq$ 80 Myr;][]{2024MNRAS.534.3022M}, and NGC 3532 \citep[300 Myr;][]{2025ApJ...995..128B}, where maximum pulsator fractions and occurrences were measured to be $\sim$80\%, $\sim$80\%, 100\%, and 63\% respectively. The higher pulsation fractions in the younger populations compared to the pulsator occurrence in the intermediate age cluster NGC 3532 suggest that the pulsation occurrence depends on age.


In this paper, we use TESS photometry to identify and characterize 487 $\delta$ Scuti stars in  20 open clusters with ages ranging from tens of Myr to $\sim$ 1 Gyr. We used this sample to investigate the relationship between pulsator occurrence, age, rotation, $T_{\rm eff}$, and metallicity.


\section{Sample Selection}

\subsection{Cluster Selection}

We selected open clusters within 500 pc. The ability to detect $\delta$ Scuti stars with TESS decreases significantly at $T$ mag $\gtrsim 11$, where noise levels become comparable to common $\delta$ Scuti pulsation amplitudes \citep[$\gtrsim$ 100 ppm;][]{2024MNRAS.528.2464R}. For the least luminous $\delta$ Scuti stars ($L \approx 10\,L_\odot$), $T\,{\rm mag}=11$ corresponds to a distance of $\sim$500 pc.



We also considered the number of stars in each open cluster, which controls the precision to which we can measure pulsator occurrence through the counting error. Our goal was to keep the counting error around $10\%$ in each cluster. Assuming a pulsator occurrence of $50\%$, which maximizes the counting error, we would need at least 25 A- and F-type stars total for the counting error to be $10\%$. Therefore, we only consider clusters with at least 25 members within the de-reddened color range in which we search for $\delta$ Scuti stars: $(G_{\rm BP}-G_{\rm RP})_0 = [-0.05, 0.6]$. 

Finally, we only searched open clusters with ages less than 1 Gyr to avoid clusters where many instability strip stars, especially the hottest $\delta$ Scuti stars, have evolved off of the main sequence. In general, there are few populous open clusters within 500 pc with ages $\geq$ 1 Gyr \citep{HuntReffert23,HuntReffert24}. In total, we found 20 open clusters that are suitable for calculating $\delta$ Scuti pulsator occurrence with our desired precision. Properties of all open clusters can be seen in Table \ref{tab:clusters}. Properties of all stars studied in this work are available in Table \ref{tab:allstars}.


\subsection{Cluster Membership}

For most open clusters, we adopt membership catalogs from \citet{HuntReffert23, HuntReffert24}, who performed a blind census of Galactic open clusters using Gaia DR3 \citep{2023Gaia} astrometry and the Hierarchical Density-Based Spatial Clustering of Applications with Noise algorithm \citep[HDBSCAN;][]{mcinnes2017hdbscan}. \citet{HuntReffert23,HuntReffert24} provide membership probabilities along with a flag identifying candidates that lie within the estimated tidal radius of each cluster, computed following \cite{1962AJ.....67..471K}.

For each cluster, we searched for $\delta$ Scuti pulsators among candidate members with this tidal-radius flag set to true. By construction, all stars within the tidal radius have membership probabilities greater than 50\%. As noted by \citet{HuntReffert24}, candidates with membership probabilities below 50\% are consistently associated with low-quality memberships.

For three open clusters, we adopted membership catalogs from \citet{liu2025revisitingopenclusters200}, who revised the \citet{HuntReffert23} catalogs to account for projection effects that can be significant for clusters within 200 pc. The clusters are NGC 2451A, $\alpha$ Per (Melotte 20), and Alessi 84, all of which exhibit substantially larger membership populations in the updated analysis compared to the original \citet{HuntReffert23} catalogs.

Three of the open clusters analyzed here have been previously studied, with their $\delta$ Scuti pulsators already identified. We therefore did not perform an independent search for the $\delta$ Scuti stars in these clusters and instead adopted the published classifications. The clusters are the Pleiades \citep{2023ApJ...946L..10B}, NGC 2516 \citep{2024A&A...686A.142L}, and NGC 3532 \citep{2025ApJ...995..128B}. Table \ref{tab:clusters} provides basic properties of all open clusters studied here.

\subsection{TESS Photometry}

We used photometry from TESS to identify $\delta$ Scuti pulsators in open clusters. We used the Full-Frame-Images (FFIs) from the extended missions, with cadences of 10-minutes and 200-seconds in the first and second extended missions, respectively. We also used 20-second and 2-minute cadence targeted observations when available.

For all but two open clusters, we used TESS light curves available from the Mikulski Archive for Space Telescopes (MAST)\footnote{https://archive.stsci.edu/}. Because FFI photometry may be reduced by multiple pipelines for a given sector and cadence, we adopt a priority scheme. We first selected FFI light curves processed by the TESS Science Processing Operations Center pipeline \citep[TESS-SPOC;][]{2020RNAAS...4..201C}. If TESS-SPOC products are not available, we instead used light curves from the Quick-Look Pipeline \citep[QLP;][]{2020RNAAS...4..204H}.

\begin{figure}
    \centering
    \includegraphics[width=\columnwidth]{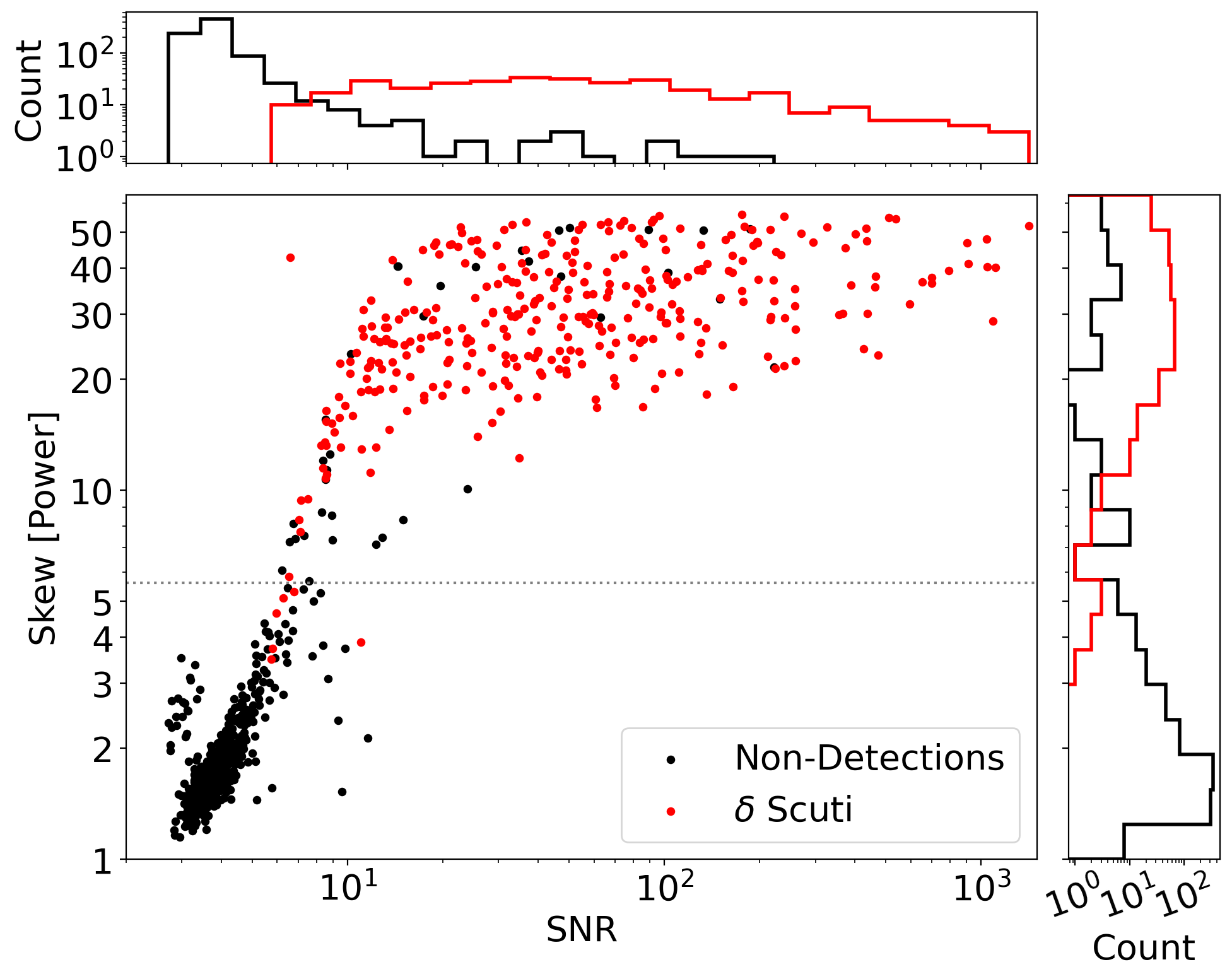}
    \caption{Scatter plot of log skewness vs. maximum SNR. Stars identified as $\delta$ Scuti pulsators are in shown as the red points, and non-detections are shown as the black points. The distributions of log skewness and SNR for $\delta$ Scuti pulsators and non-detections are shown along each axis, with the same color scheme as the scatter plot. The gray dotted line marks log skewness = 0.75, which is the lower limit used when conducting the injection and recovery tests described in \S\ref{subsec:i+r}. This plot does not include stars in the Pleiades, NGC 2516, and NGC 3532, where the $\delta$ Scuti stars were identified in previous literature \citep[][respectively]{2023ApJ...946L..10B,2024A&A...686A.142L,2025ApJ...995..128B}.}
    \label{fig:skew_v_snr}
\end{figure}

The light curves were downloaded and processed using the \textsc{LightKurve} Python package \citep{2018ascl.soft12013L}. We performed outlier rejection, normalization, NaN removal, and flattening of all light curves using the methods provided in \textsc{LightKurve} \citep{2018ascl.soft12013L}.

We used TESS-Gaia Light Curves \citep[TGLCs;][]{2023AJ....165...71H} for the open clusters Stock 2 and NGC 3532. Specifically, we used the same light curves for NGC 3532 as those used by \citet{2025ApJ...995..128B} to mitigate the effects of blending in the crowded field around NGC 3532. We generated our own TGLCs for Stock 2 members, as reduced light curves for a large fraction of Stock 2 members were not readily available on MAST. To do this, we used the \texttt{quick\textunderscore lc.tglc\textunderscore lc} method from the \textsc{tglc} Python package. For every star searched, light curves from each unique sector and cadence pair were stitched together into one combined light curve. 

We note that the PSF-based Approach to TESS High quality data Of Stellar clusters database \citep[PATHOS;][]{2021MNRAS.505.3767N} provides high-precision light curves for open cluster members from TESS FFIs. However, PATHOS only covers up to TESS Sector 26, and therefore does not include the extended mission sectors required for this study of $\delta$ Scuti stars.

We calculated amplitude spectra for all individual sectors of photometry, as well as for all sector-combined light curves using the \texttt{LombScargle} method from the \textsc{AstroPy} Python package \citep{astropy:2013,astropy:2018, astropy:2022}. Ultimately, we used the sector-combined amplitude spectra to search for $\delta$ Scuti pulsators. The amplitude spectra for all $\delta$ Scuti stars found in this work are shown in Figure \ref{fig:stacked_amp_spec}.

\subsection{Contamination/Blending} \label{subsec:blend}

Blending is expected for some sources because of the large 21$\arcsec,$ TESS pixels. However, the high amplitudes of $\delta$ Scuti pulsations, together with the fact that the brightest stars in the open cluster fields are cluster members, make contamination from fainter stars negligible. Furthermore, few astrophysical signals resemble $\delta$ Scuti pulsations. The primary source of contamination therefore arises from blended A- or F-type cluster members, where one or both stars are pulsating, making it difficult to determine which pulsation modes originate from each star. To identify such cases, we calculated the nearest-neighbor distance for every A/F-type cluster member and visually compared the amplitude spectra of stellar pairs separated by less than 10 TESS pixels. Based on these comparisons, we classified each star as either a pulsator or a non-pulsator.


\subsection{Photometric Quality Cuts}

For each cluster, we automatically discarded poor-quality photometry on a star-by-star, sector-by-sector basis prior to constructing sector-combined light curves. We estimated the noise level in each amplitude spectrum by measuring the mean amplitude within a quiet, high-frequency region within 95\% of the Nyquist frequency. This was then compared to the expected noise level derived from simulated light curves and amplitude spectra, generated using photon noise estimates from the \textsc{Ticgen} Python package \citep{jaffe_barclay_2017, 2018AJ....156..102S}.

For each star and sector, we computed the ratio of the measured noise to the expected noise. Photometry with noise ratios exceeding the 95th percentile of all stars and sectors was discarded. This procedure is performed independently for each cadence.

\subsection{Interstellar Reddening \& Extinction}

For each cluster, we de-reddened the Gaia $G_{\rm BP}-G_{\rm RP}$ using the median $A_V$ values provided in the cluster member catalogs from \citep{HuntReffert23,HuntReffert24}. We find $E(B-V)$ assuming an extinction law with $R_V=3.1$. We then estimate $E(G_{\rm BP}-G_{\rm RP}) = E(B-V)/0.76$, found from Table 2 of \cite{2019ApJ...877..116W}. De-reddening is performed before any color constraints are applied to the cluster catalogs. We also corrected for extinction using the same $A_V$ values.

\section{Calculating Pulsator Occurrence} \label{sec:occurrence}

\begin{figure*}
    \centering
    \includegraphics[width=\linewidth]{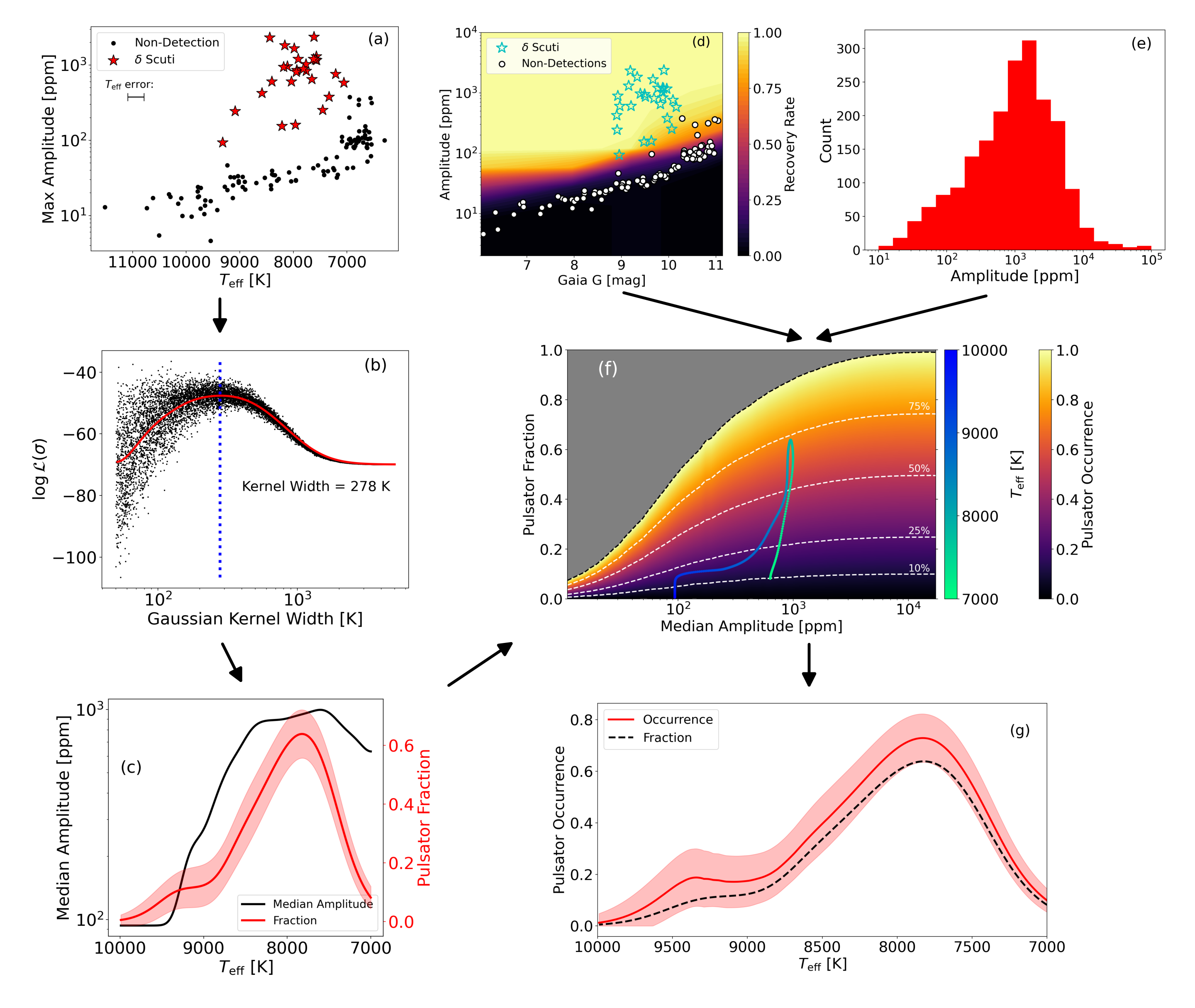}
    \caption{Flowchart illustrating the procedures described in \S\ref{sec:occurrence} for the open cluster NGC 6475. Panel (a) shows the maximum amplitude vs. $T_{\rm eff}$. Red stars indicate $\delta$ Scuti pulsators, black points show non-detections. Panel (b) shows the log-likelihood of the Bernoulli distribution (i.e. Equation \ref{eq:ll}) as a function of the Gaussian kernel width from a Monte Carlo simulation. The red curve is the binned and smoothed log-likelihood, and the blue dotted line indicates the kernel width that maximizes the log-likelihood. Panel (c) shows the median amplitude and pulsator fraction as functions of $T_{\rm eff}$, using the optimized kernel width. Panel (d) shows $\delta$ Scuti recovery rates as a function of amplitude and apparent Gaia $G$ magnitude. Cyan stars show $\delta$ Scuti stars, white circles show non-detections. The background colors show recovery rate. Panel (e) shows the $\delta$ Scuti amplitude distribution for all observed $\delta$ Scuti stars from Kepler \citep{2019MNRAS.485.2380M}. Panel (f) shows the pulsator occurrence as a function of pulsator fraction and median amplitude. The inferno color-map shows the pulsator occurrence. The median amplitudes and pulsator fractions from panel (c) are over-plotted as the cyan and blue curve, which tracks $T_{\rm eff}$. Panel (g) shows the resulting pulsator occurrence as a function of $T_{\rm eff}$ as the red curve. Pulsator fraction is also shown again as the black dashed curve. The red shaded region shows the counting error.}
    \label{fig:summary_m7}
\end{figure*}

\subsection{$\delta$ Scuti Identification} \label{subsec:identification}

The first step to calculate the pulsator occurrence in an open cluster is to identify $\delta$ Scuti pulsators. We followed a similar method from \citet{2019MNRAS.485.2380M}, by using the skewness of the distribution of peak amplitudes, which will be systematically larger for $\delta$ Scuti stars than for stars with no high-frequency variability. This is slightly different from the method used by \citet{2019MNRAS.485.2380M}, who calculated the skewness using all Fourier amplitudes. In previous literature, a fixed lower frequency boundary \citep[such as 5 ${\rm d}^{-1}$;][]{2019MNRAS.485.2380M, 2024ApJ...972..137G} was set in order to avoid false positives from sources of rotational modulation or $\gamma$ Doradus (Dor) pulsations \citep{Kaye_1999}. For this work, we used a variable lower frequency boundary, set by the empirical $\delta$ Scuti period-luminosity relation from \citet{2022MNRAS.516.2080B} (their equation 4), which is thought to follow the frequency of the fundamental mode.

However, we also confirmed all stars that could be $\delta$ Scuti pulsators by visually inspecting the amplitude spectra of stars with log skewness $\geq$ 0.4. A Gaia CMD and HR diagram showing all $\delta$ Scuti stars and non-detections is shown in Figure \ref{fig:CMD_HR}. The distributions of log skewness and maximum signal-to-noise ratio (SNR) for all observed stars are shown in Figure \ref{fig:skew_v_snr}.




To validate that we properly identified the $\delta$ Scuti pulsators in an open cluster, we plot the maximum amplitude vs. $T_{\rm eff}$. Pulsators typically have a maximum amplitude of order unity mmag. We did not correct the maximum amplitudes for apodization, whereas \citet{2025MNRAS.542.2866M} applied such a correction. The non-pulsators should form a ``non-detection ridge'', presumably due to white noise, which increases with cooler, dimmer stars in an open cluster. This plot for the open cluster NGC 6475 (Messier 7) is shown in panel (a) of Figure \ref{fig:summary_m7}.


\subsection{Measuring Pulsator Fraction Over $T_{\rm eff}$} \label{sec:frac}

The next step involved measuring the pulsator fraction as a continuous function of $T_{\rm eff}$. In previous work, the pulsator fraction across the instability strip was measured discretely by binning in apparent magnitude or color \citep[e.g.][]{2023ApJ...946L..10B, 2024ApJ...972..137G, 2024MNRAS.528.2464R, 2024MNRAS.534.3022M,2025MNRAS.542.2866M, 2025ApJ...995..128B}. However, given that we now aim to compare pulsator occurrences across multiple different stellar populations, it is reasonable to work with fundamental stellar parameters such as $T_{\rm eff}$, rather than photometric parameters, as was done for Kepler by \citet{2019MNRAS.485.2380M}. We used $T_{\rm eff}$ values from the TESS Input Catalog \citep[TIC;][]{2019AJ....158..138S}.

\begin{figure}
    \centering
    \includegraphics[width=\columnwidth]{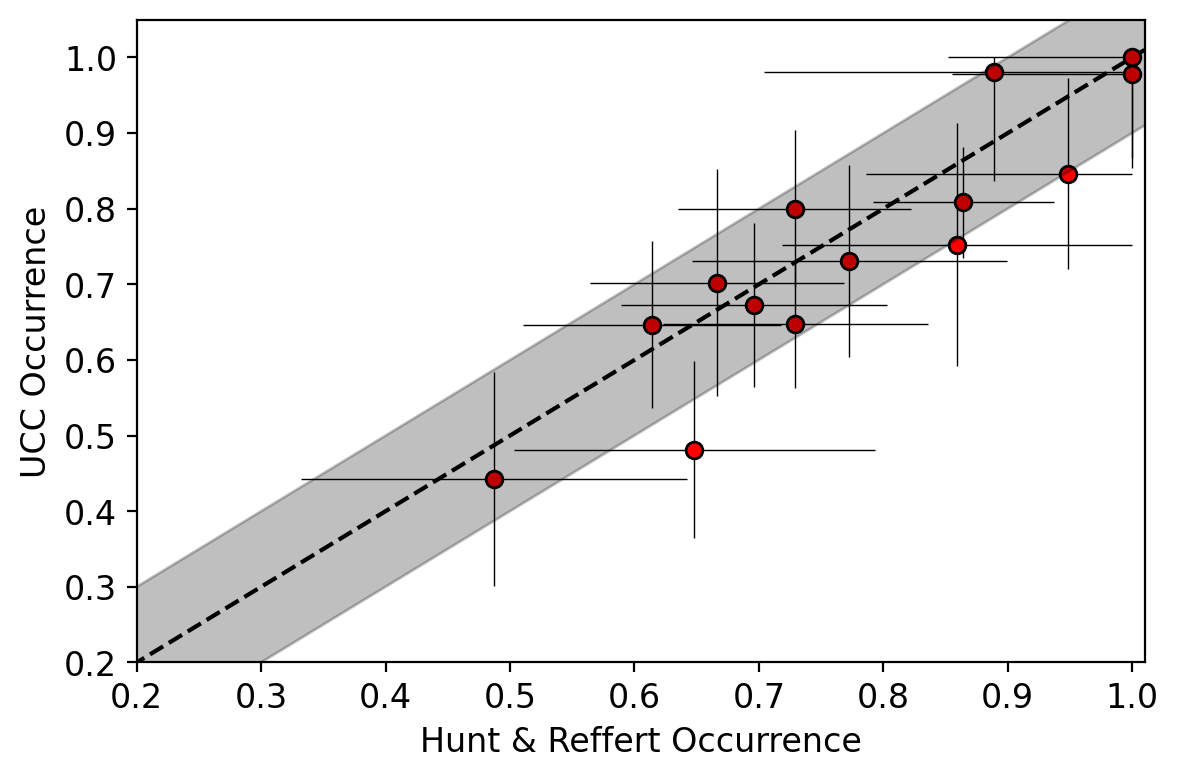}
    \caption{The pulsator occurrence rate calculated from UCC membership \citep{2023MNRAS.526.4107P} vs. that found from the \citet{HuntReffert23, HuntReffert24} catalog. Each red point shows one open cluster. The black dashed line shows the 1:1 ratio, and the shaded regions show the $\pm10\%$ uncertainty.}
    \label{fig:hr_v_ucc}
\end{figure}

By observing $\delta$ Scuti pulsators separated into coeval, equal-metallicity populations, we can study how pulsator occurrence changes as a function of age, metallicity, and $T_{\rm eff}$. Rather than discretely binning over color or $T_{\rm eff}$, we measured pulsator occurrence as a continuous function of $T_{\rm eff}$, avoiding ambiguities associated with binning. We estimated pulsator fraction as a function of $T_{\rm eff}$ using a Gaussian kernel given by:

\begin{equation} \label{eq:kernel}
    w_i(T_{\rm eff}, \sigma) = \frac{1}{\sqrt{2\pi}\sigma}\exp\biggl[\frac{-(T_{\rm eff} - T_{{\rm eff,}\,i})^2}{2\sigma^2}\biggr]\, ,
\end{equation}

\noindent where $T_{\rm eff}$ is the effective temperature where the fraction is to be evaluated, and $T_{{\rm eff},i}$ is the effective temperature of one star in the sample. The kernel width is given by $\sigma$. The pulsator fraction, $f$, for a given $T_{\rm eff}$ and $\sigma$ was found by:

\begin{equation} \label{eq:frac}
    f(T_{\rm eff}, \sigma) = \frac{\sum_{i=1}^N p_i\, w_i(T_{\rm eff}, \sigma) }{\sum_{i=1}^N w_i(T_{\rm eff}, \sigma)}\, ,
\end{equation}

\noindent where $p_i$ is a boolean flag ($p_i=1$ for $\delta$ Scuti pulsators, $p_i=0$ for non-detections), and $N$ is the sample size. The pulsator fraction $f$ is evaluated at every $T_{\rm eff}$ to obtain a continuous function of $T_{\rm eff}$. We computed $f$ over $T_{\rm eff} = 7000$–$10,000$ K. Note that the fraction is unweighted by the total number of stars.

\begin{figure*}
    \centering
    \includegraphics[width=\linewidth]{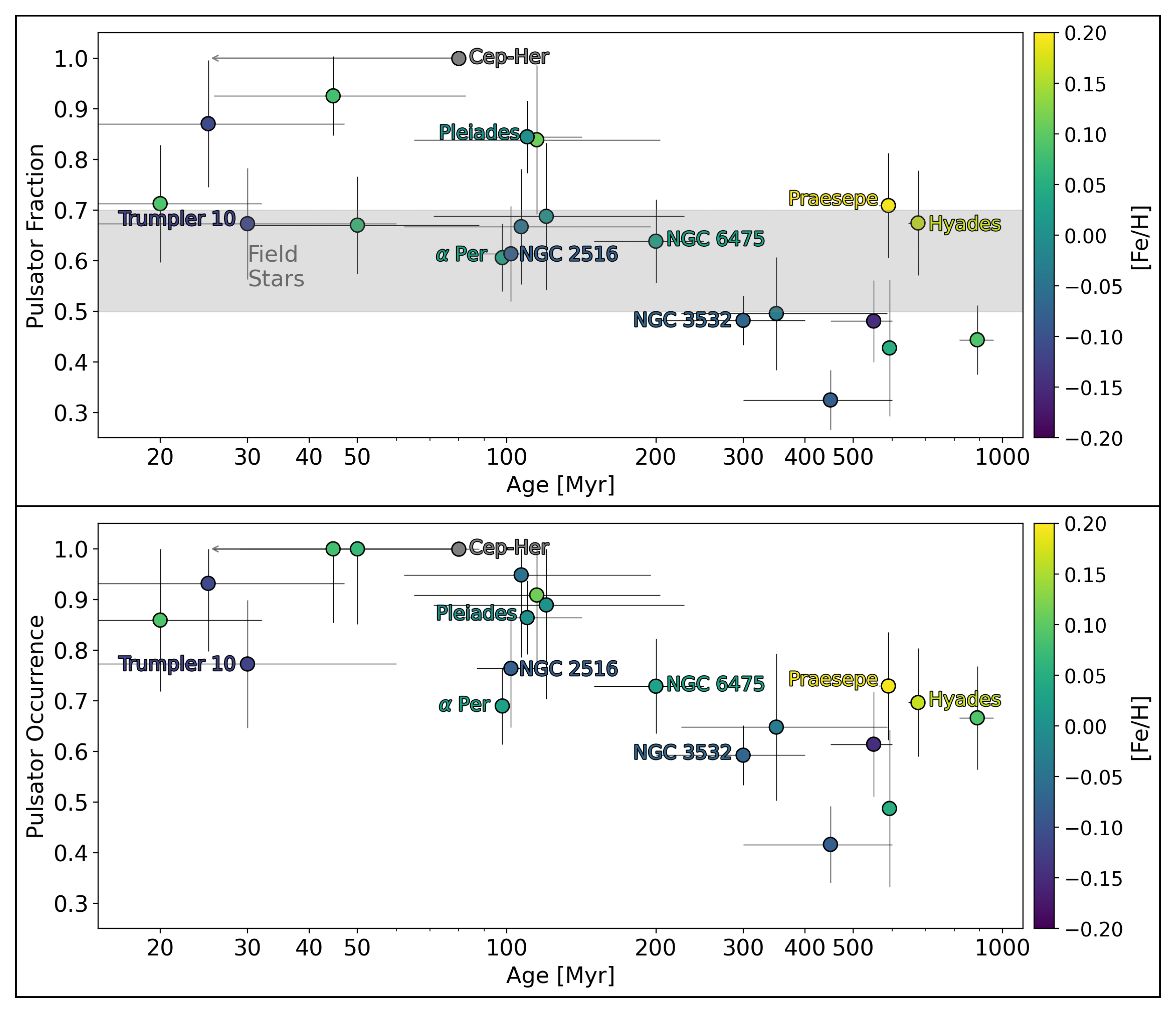}
    \caption{Pulsator fraction (top panel) and pulsator occurrence (bottom panel) as a function of cluster age. Colors show the [Fe/H] of each cluster from previous literature (see Table \ref{tab:clusters}). The gray point with the arrow shows the Cep-Her complex, which is made up of stellar populations with ages between 25 and 80 Myr \citep{2024MNRAS.534.3022M}. The gray band in the top panel shows the pulsator fraction measured among field $\delta$ Scuti stars \citep[e.g.][]{2019MNRAS.485.2380M,2024MNRAS.528.2464R,2025MNRAS.542.2866M}.}
    \label{fig:occurrence_fraction}
\end{figure*}

The fraction is highly dependent on our choice of $\sigma$. Smaller $\sigma$ will lead to a narrower, taller distribution, and vice versa for a larger $\sigma$, which will strongly influence our results. Therefore, it was crucial that we chose $\sigma$ per cluster in a non-biased, unambiguous way, as we expected that the distribution of pulsators and non-pulsators in $T_{\rm eff}$ would change from cluster to cluster.

We optimized for $\sigma$ by maximizing the likelihood. In this case, we maximized the likelihood of a Bernoulli distribution, given the boolean nature of this problem. Specifically, we maximized:

\begin{equation} \label{eq:ll}
    \ln\mathcal{L}(\sigma) = \sum_{i=1}^N p_i\ln\hat{f}_i(T_{{\rm eff}}, \sigma)+(1-p_i)\ln[1-\hat{f}_i(T_{{\rm eff}}, \sigma)] \, ,
\end{equation}

\noindent where $\hat{f}$ is Equation \ref{eq:frac} found via ``leave-one-out'' (LOO) cross-validation, which we performed such that each star is predicted as either a pulsator or non-pulsator based only on neighboring stars. In other words, $\hat{f}$ is Equation \ref{eq:frac} calculated excluding the $i$-th star, and stops $\sigma$ from collapsing towards zero. $\hat{f}$ is given by:

\begin{equation}
    \hat{f_i} = \frac{\sum_{j\neq i}^Np_j\,w_j(T_{\rm eff},\sigma)}{\sum_{j\neq i}^Nw_j(T_{\rm eff},\sigma)}\,.
\end{equation}


We calculated Equation \ref{eq:ll} in a Monte Carlo simulation with 10,000 trials. In each trial, we varied the $T_{\rm eff}$ of each star within the uncertainties, assuming that they are normally distributed. We randomly drew values for $\sigma$ from a log-uniform distribution with values from 50-5000 K. After all trials, we binned and smoothed the likelihood, and find the $\sigma$ corresponding to the maximum likelihood. This $\sigma$ was then used to calculate Equations \ref{eq:kernel} and \ref{eq:frac}. In most open clusters, $\sigma$ was on the order of hundreds of Kelvin, with an average kernel width of $\approx 360$ K across all 20 open clusters. An example of this process in the open cluster NGC 6475 is shown in panel (b) of Figure \ref{fig:summary_m7}.

The uncertainty on the pulsator fraction, $\sigma_f$, at each $T_{\rm eff}$ can be assumed to be the counting error:

\begin{equation}
    \sigma_{f} = \sqrt{\frac{f(1-f)}{N_{\rm eff}}}\,,
\end{equation}

\noindent where $f$ is the pulsator fraction found via Equation \ref{eq:frac} and $N_{\rm eff}$ is the effective sample size, found by:

\begin{equation}
    N_{\rm eff} = \frac{[\sum_i^Nw_i(T_{\rm eff},\sigma)]^2}{\sum_i^Nw_i^2(T_{\rm eff},\sigma)}\,,
\end{equation}

\noindent where $w_i(T_{\rm eff},\sigma)$ is found via Equation \ref{eq:kernel}. These errors are propagated when calculating the pulsator occurrence.

Pulsator occurrence also depends on the median $\delta$ Scuti pulsation amplitude. We estimate the median amplitude as a continuous function of $T_{\rm eff}$ using a Gaussian kernel with the same width as for the pulsator fraction. Panel (c) of Figure \ref{fig:summary_m7} shows both quantities versus $T_{\rm eff}$ for NGC 6475, which we compare to assess how occurrence varies with temperature.



\subsection{Injection and Recovery Tests} \label{subsec:i+r}

Following \citet{2025ApJ...995..128B}, we injected sinusoids into the light curves of the non-detections with amplitudes ranging between 1 and $10^6$ ppm, and random frequencies and phase offsets. We calculated the amplitude spectrum of each non-detection with the injected sinusoid, and a recovery was considered when the log skewness was $\geq 0.75$. The recovery rate was found by binning over amplitude and apparent magnitude and calculating the fraction of recoveries in each bin. An example of this process for the open cluster NGC 6475 is shown in panel (d) of Figure \ref{fig:summary_m7}.

\subsection{Calculating Pulsator Occurrence}


In \citet{2025ApJ...995..128B}, stars were randomly assigned as pulsators or non-pulsators, with the number of pulsators determined by the chosen pulsator occurrence, set between 0 and 1. Stars classified as pulsators were assigned one random pulsation amplitude drawn from the amplitude distribution found from field $\delta$ Scuti stars observed by Kepler \citep[panel (e) of Figure \ref{fig:summary_m7};][]{2019MNRAS.485.2380M}. The recovery rate for each pulsator was then determined from its apparent magnitude and assigned amplitude, using the injection and recovery tests described in \S\ref{subsec:i+r}. The number of recoveries was the sum of the recovery rates, and a pulsator fraction was measured as the number of recoveries over the total sample size. To account for different median amplitudes, this process was repeated for different amplitude distributions, which was handled by shifting the Kepler distribution uniformly in log-space to match the desired median amplitude. The plot resulting from this exercise (i.e. the $\delta$ Scuti occurrence plot) can be seen in panel (f) of Figure \ref{fig:summary_m7}, for the open cluster NGC 6475. 

The occurrence plot, combined with the functions of median amplitude and pulsator fraction described in \S\ref{sec:frac}, allows us to find pulsator occurrence as a continuous function over $T_{\rm eff}$ for any open cluster. The resulting pulsator occurrence as a function of $T_{\rm eff}$ for the open cluster NGC 6475 is shown in panel (g) of Figure \ref{fig:summary_m7}, along with the fraction for comparison. Note that the pulsator occurrence will always be greater than the pulsator fraction.

We note that the occurrence rates are largely insensitive to the choice of open cluster membership catalog. We calculated occurrence rates with memberships from the Unified Cluster Catalog \citep[UCC;][]{2023MNRAS.526.4107P} for a subset of our clusters, which yields occurrence rates that differ from those based on \citet{HuntReffert23, HuntReffert24} by less than our target uncertainty of $\pm10\%$. These results are shown in Figure \ref{fig:hr_v_ucc}. Thus, our conclusions are unchanged regardless of the catalog adopted.

\subsection{Effect of equal-mass Binaries}

When calculating pulsator fractions and occurrences, we count each unresolved source only once. Consequently, unresolved equal-mass binaries may bias the inferred pulsator fractions and occurrences. To estimate the magnitude of this effect, we express the corrected pulsator fraction in terms of the measured fraction $f$ from Equation \ref{eq:frac}, the equal-mass binary fraction $b$, and the fraction of unresolved companions that are pulsators, $g$:

\begin{equation}
f_{\rm scaled} = \frac{f+bg}{1+b}.
\end{equation}

In the worst case scenario of an equal-mass binary fraction of $b=10\%$ and no pulsating companions ($g=0$), the inferred pulsator fraction is reduced by $\sim9\%$, comparable to our adopted uncertainty ($\sim10\%$). If the pulsator fraction among unresolved companions is the same as the primary stars ($g=f$), then $f_{\rm scaled}=f$, making the inferred pulsator fraction independent of the equal-mass binary fraction. We consider $b=10\%$ as the worst case scenario, as \citet{2018MNRAS.474.4322M} showed that the fraction of A/F-type primaries with $0.9\leq q \leq 1.0$ is about 6\%. This corresponds to an equal-mass binary fraction of $\approx4\%$ if we assume a multiplicity fraction of 70\%.

In equal-mass binaries, flux dilution doubles the observed flux, potentially affecting the detection of $\delta$ Scuti stars. However, the associated increase in photon noise is only a factor of $\sqrt{2}$, which is negligible because the noise levels remain orders of magnitude below the typical $\sim$mmag pulsation amplitudes.

\section{Pulsator Occurrence, Age, Metallicity, $T_{\rm eff}$, and Rotation}

\subsection{Pulsator Occurrence and Age}

The lower panel of Figure \ref{fig:occurrence_fraction} shows the maximum pulsator occurrence in all 20 open clusters as a function of cluster age.
For clusters $<$200 Myr old, we can see that the pulsator occurrence is consistently higher than the pulsator fraction measured in field stars samples \citep[50-70\%;][]{2019MNRAS.485.2380M,2024MNRAS.528.2464R, 2025MNRAS.542.2866M}. 

Beyond an age of $\sim 200$ Myr, the pulsator occurrence decreases. The average pulsator occurrence in open clusters older than 200 Myr is 62$\pm3$\% . Meanwhile, the average pulsator occurrence in open clusters younger than 200 Myr is 88$\pm3$\%. Therefore, the difference in the pulsator occurrence between young and old open clusters is statistically significant, and shows that the $\delta$ Scuti pulsator occurrence decreases with age.


\begin{figure}
    \centering
    \includegraphics[width=\columnwidth]{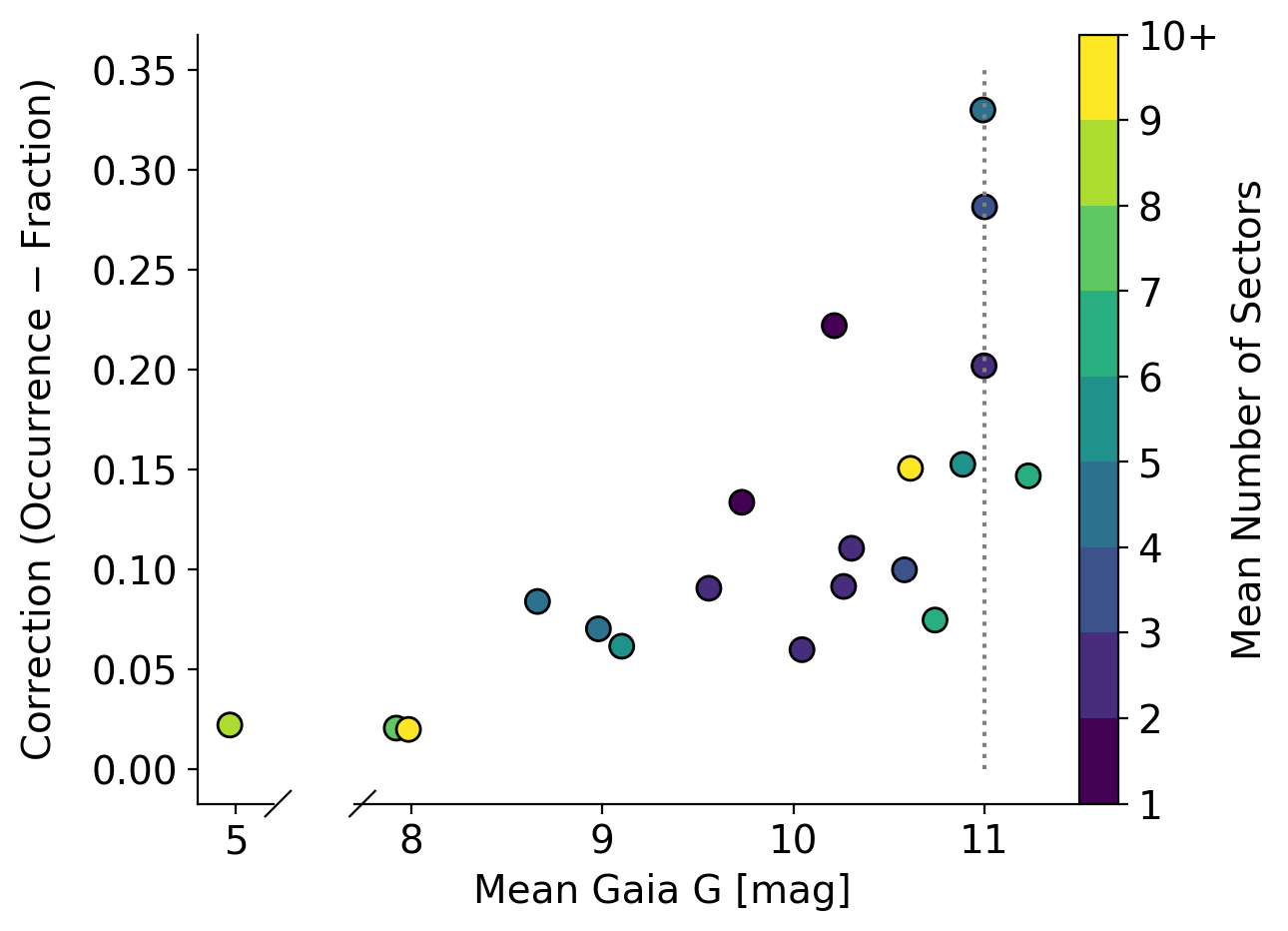}
    \caption{The occurrence correction (defined as the difference between the maximum occurrence and maximum fraction) as a function of the mean Gaia $G$ magnitude in each cluster. Colors show the average number of TESS sectors available for each open cluster.}
    \label{fig:correction}
\end{figure}



When viewing pulsator fraction vs. age, the trend is less clear (Figure \ref{fig:occurrence_fraction}), as the fraction among young open clusters is not as consistently high as is seen with the pulsator occurrence. The average pulsator fraction for clusters younger than 200 Myr is $74\pm4\%$, and while the average for clusters older than 200 Myr is $52\pm4\%$. This shows the importance of applying the occurrence correction to the fraction for certain open clusters.


Figure \ref{fig:correction} shows the occurrence correction  (i.e. the difference between the maximum occurrence and maximum fraction) as function of the mean Gaia $G$ magnitude for each cluster. As expected, we see that larger corrections are warranted when observing fainter samples. We see that the correction can increase rapidly from Gaia $G\geq 10$, which coincides with the detection limit of TESS for $\delta$ Scuti stars \citep{2024MNRAS.528.2464R}. Furthermore, there is no apparent change in the correction needed between the 8th magnitude clusters (the Pleiades and Praesepe) and the 5th magnitude cluster (the Hyades). This indicates that no completeness correction is necessary for samples brighter than 8th magnitude, and $\delta$ Scuti samples in this regime can be treated as effectively complete, in agreement with \citet{2024MNRAS.528.2464R}.

\begin{figure}
    \centering
    \includegraphics[width=\columnwidth]{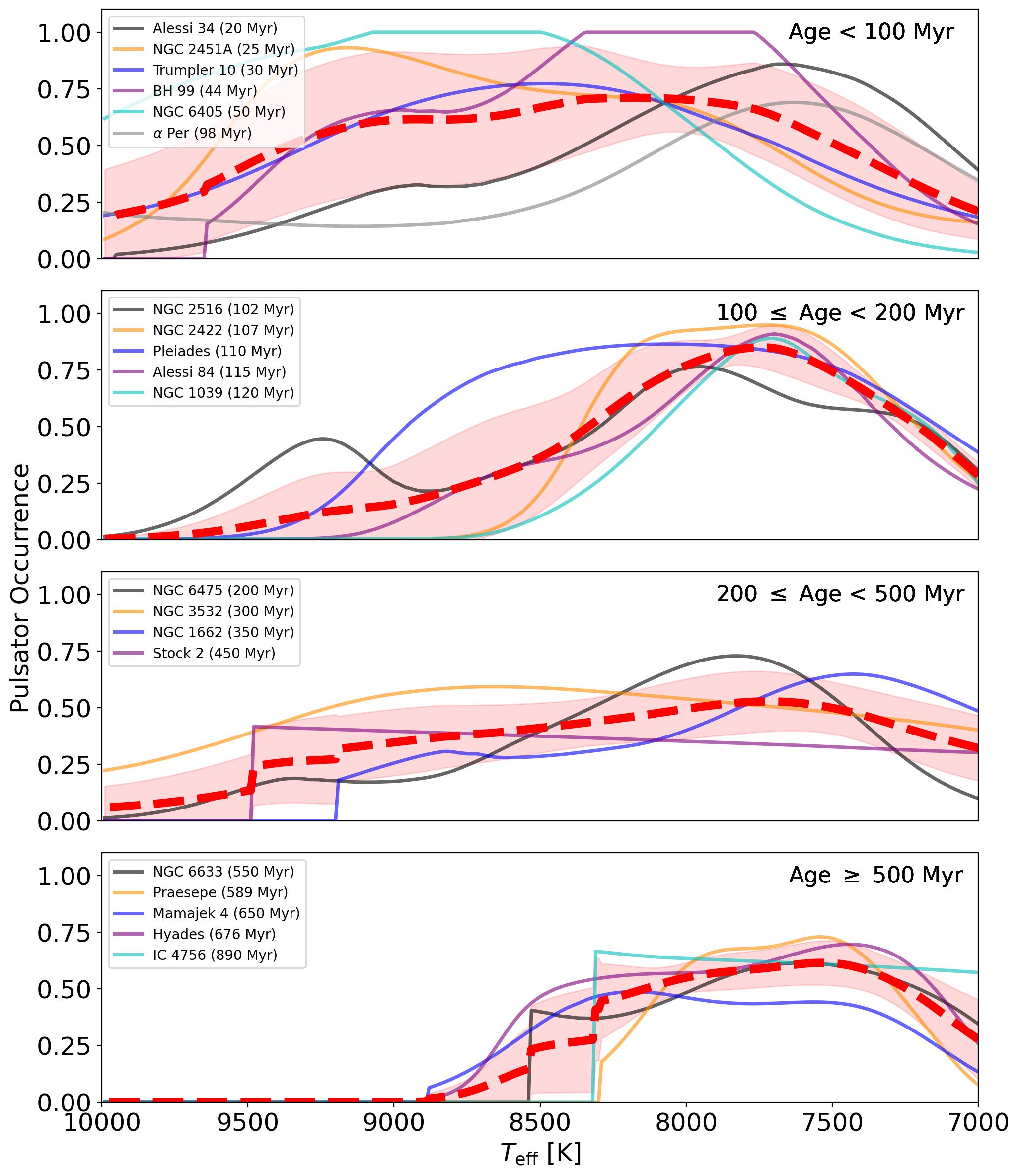}
    \caption{Pulsator occurrence vs. $T_{\rm eff}$ for all open clusters, separated and sorted by age, with the age range listed in each panel. The red dashed curve shows the average pulsator occurrence over $T_{\rm eff}$, and the red shaded region shows the $1\sigma$ deviation.}
    \label{fig:occur_v_teff_v_age}
\end{figure}

\subsection{Pulsator Occurrence and Metallicity}

The pulsator occurrence rates in some older clusters are comparable to the pulsator occurrences seen in some clusters less than 100 Myr old. One of these clusters is the $\sim$600 Myr old Praesepe, which is the most metal-rich cluster studied in this work \citep{2018ApJ...863...67G, 2020A&A...633A..38D}, which suggests that more metal-rich stars may be able to continue pulsating for longer periods of time. The $\sim 700$ Myr old Hyades also shows an elevated pulsator occurrence and is also relatively metal-rich \citep{2016MNRAS.457.3934L,2018ApJ...863...67G}. However, most clusters studied here have [Fe/H] close to Solar, so additional metal-rich and metal-poor open clusters will need to be studied in the future to solidify any existing trend between pulsator occurrence and metallicity.

\subsection{Pulsator Occurrence and $T_{\rm eff}$}

Although the maximum overall pulsator occurrence in each cluster is a useful global statistic, $\delta$ Scuti stars can exist over a $\sim2000$ K range of temperatures, so it is important to also examine how the distribution of $\delta$ Scuti pulsators in $T_{\rm eff}$ changes with age. Figure \ref{fig:occur_v_teff_v_age} shows the pulsator occurrence vs. $T_{\rm eff}$ for all open clusters studied here, grouped in different age bins, with the mean pulsator occurrence shown as the red dashed curves. In accordance with Figure \ref{fig:occurrence_fraction}, we can see that the peaks of the functions decrease at older ages. However, we can now see that the pulsator occurrences in hotter $\delta$ Scuti stars ($\gtrsim$ 8500 K) appear to be larger in younger open clusters. This suggests that hotter $\delta$ Scuti stars may stop pulsating sooner than their cooler counterparts.

\subsection{Pulsator Occurrence and Rotation}

Previous work has shown that rotation also affects pulsator occurrence, where faster rotators are more likely to pulsate \citep{2023ApJ...946L..10B,2024ApJ...972..137G,2024MNRAS.534.3022M,2025ApJ...995..128B}. Therefore, it is important to examine how age and rotation together affect the pulsator occurrence.

\begin{figure}
    \centering
    \includegraphics[width=\columnwidth]{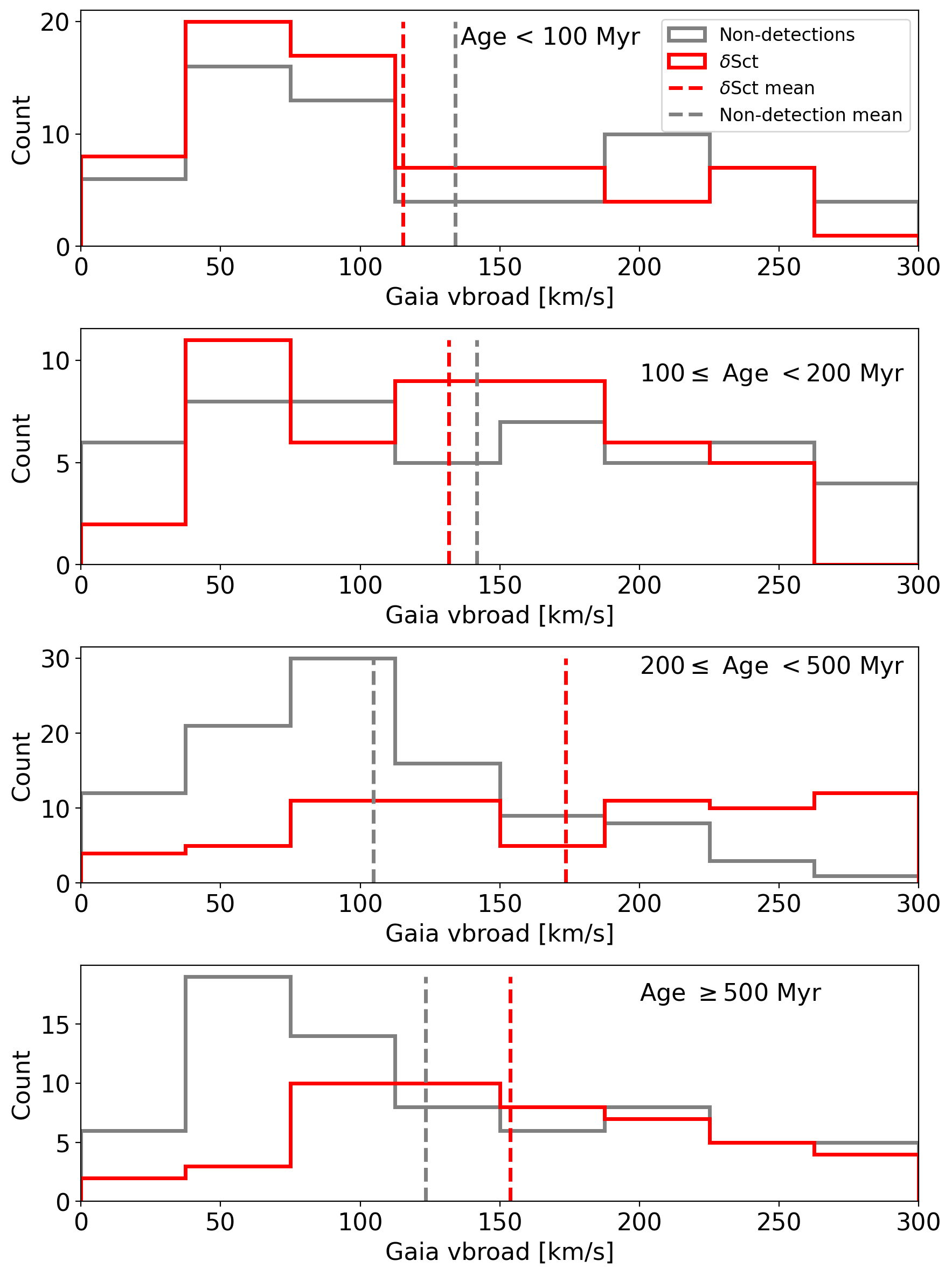}
    \caption{Gaia \texttt{vbroad} histograms of the $\delta$ Scuti pulsators (red) and non-detections (gray) binned by age, with the youngest stars on top. The red and gray dashed lines show the mean the distributions for the $\delta$ Scuti pulsators and non-detections, respectively. These histograms only include stars within the empirical bounds of the instability strip from \citet{2019MNRAS.485.2380M}.}
    \label{fig:vbroad_hists}
\end{figure}

We use Gaia \texttt{vbroad} as our proxy for rotation, which is an estimation of the projected rotational velocity $v\sin i$ from the Gaia RVS spectrograph \citep{2018A&A...616A...5C, 2023A&A...674A...8F}. \citet{2024MNRAS.534.3022M} compared the Gaia \texttt{vbroad} to cataloged $v\sin i$ values from \citet{2012A&A...537A.120Z} and found that \texttt{vbroad} performs well for velocities between 75 and 200 ${\rm km\,s^{-1}}$, but underestimates the $v\sin i$ of slow rotators ($v\sin i\lesssim 50\,{\rm km\,s^{-1}}$), and overestimates the $v\sin i$ of rapid rotators ($v\sin i\gtrsim 200\,{\rm km\,s^{-1}}$). We note that we use the Gaia \texttt{vbroad} for all stars, excluding 49 stars in the Pleiades, where $v\sin i$ was measured using spectra from the Center for Astrophysics survey \citep{2020ApJ...901...91T,2021ApJ...921..117T,2023ApJ...946L..10B}.

Figure \ref{fig:vbroad_hists} shows the Gaia \texttt{vbroad} distributions for the $\delta$ Scuti pulsators and non-detections. On average, $\delta$ Scuti pulsators in older clusters rotate more rapidly than those in younger clusters. Together, these trends indicate that older stellar populations have lower overall $\delta$ Scuti occurrence, with the remaining pulsators likely persisting due to their rapid rotation.

\begin{figure}
    \centering
    \includegraphics[width=\columnwidth]{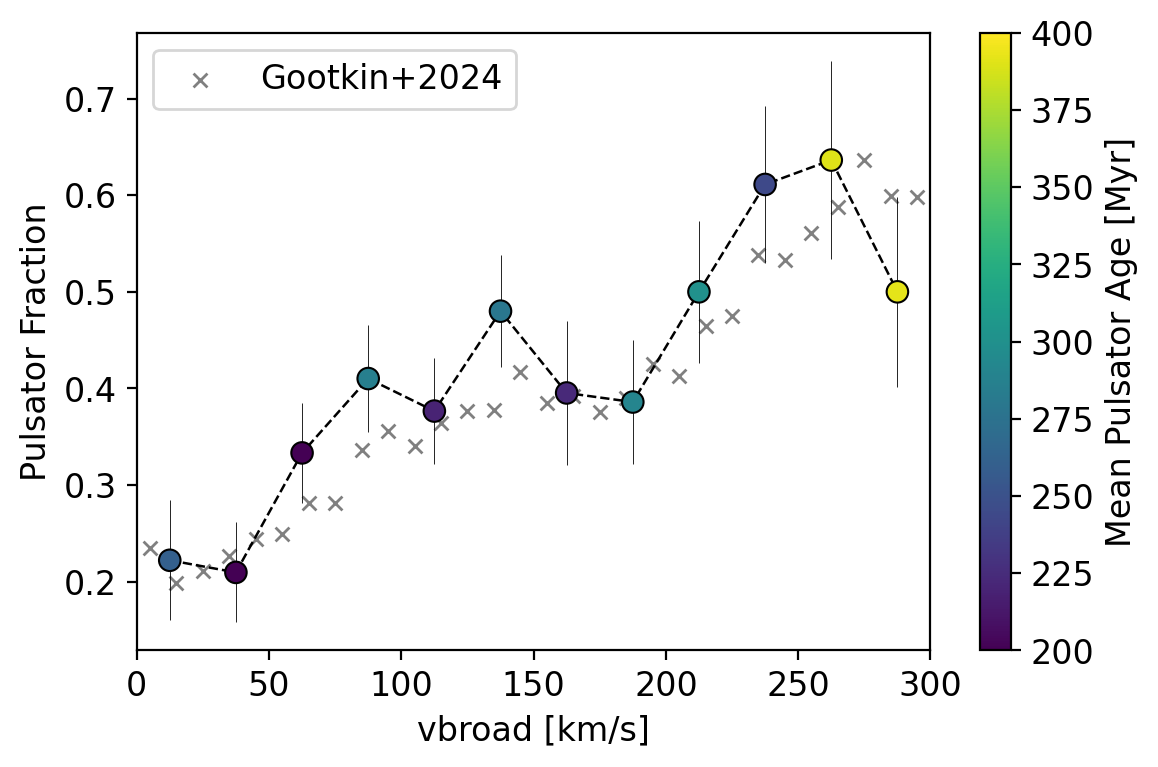}
    \caption{Pulsator fraction vs. Gaia \texttt{vbroad}, with colors showing the mean $\delta$ Scuti pulsator age in each \texttt{vbroad} bin. Results from \citet{2024ApJ...972..137G} are shown as the gray crosses.}
    \label{fig:frac_v_vbroad}
\end{figure}

We plot the pulsator fraction vs. \texttt{vbroad} in a similar way to \citet{2024ApJ...972..137G} in Figure \ref{fig:frac_v_vbroad}. We binned all stars studied here by \texttt{vbroad}, regardless of the cluster to which they belong. As a consequence, we can only measure fraction, since pulsator occurrence can only be inferred in coeval, equidistant populations. In addition, we only considered stars brighter and bluer than the empirical red edge found by \citet{2024ApJ...972..137G}. Our results agree very well with those found by \citet{2024ApJ...972..137G} (gray crosses in Figure \ref{fig:frac_v_vbroad}), in that we see a considerable increase in pulsator fraction with faster rotators. Since we have age information available, we calculated the mean $\delta$ Scuti pulsator age within each \texttt{vbroad} bin. We find a similar result as in Figure \ref{fig:vbroad_hists}; that on average, more slowly rotating pulsators tend to be younger, and older pulsators tend to be more rapidly rotating. This is the opposite of the trend expected in the general stellar population, where stars typically spin-down as they age due to expansion.

We note that \citet{2025ApJ...978...53W} reported an apparent decrease in the rotation velocity of $\delta$ Scuti stars with increasing age, based on a sample of field $\delta$ Scuti stars observed with LAMOST and TESS. This trend is opposite to what we find here. One possible source of this discrepancy is the difference in sample selection. We analyze stars in open clusters, whereas \citet{2025ApJ...978...53W} consider field stars, which are known to exhibit different $v\sin i$ and rotation axis inclination distributions \citep{2006ApJ...648..580H,2018A&A...612L...2K}.

\subsection{Physical Interpretation}

$\delta$ Scuti stars pulsate because of a partial helium ionization zone near the surface, which drives pulsations due to periodic changes in opacity \citep{2004A&A...414L..17D}. If this ionization zone is sufficiently depleted of helium, then $\delta$ Scuti pulsations may either have amplitudes too low to be detected, or stop altogether.

It is well known that helium can gravitationally diffuse into deeper layers of stars over time \citep{1973A&A....23..221B}. Therefore, it is possible that helium can settle out of the near-surface ionization zone of $\delta$ Scuti stars as they age. Models of helium settling in A-type stars \citep[e.g.][]{2005A&A...443..627T, 2020A&A...633A..23D} show that helium diffusion occurs quickly relative to the main sequence lifetimes of these stars. Specifically, 50\% of helium diffusion can occur by an age of 100 Myr and 80\% depletion by 500 Myr. It is still not clear exactly how much depletion would cause a $\delta$ Scuti star to stop pulsating. Figure \ref{fig:occurrence_fraction} shows a noticeable decrease in the pulsator occurrence in clusters much older than 100 Myr, suggesting that $\delta$ Scuti stars may stop pulsating once the helium depletion in the ionization zone exceeds 50\%.

\begin{deluxetable*}{llll} \label{tab:allstars}
\caption{Properties of all stars searched in this work. This table is available in machine-readable format.}
\tablehead{
Column name & Header & Unit & Description
}
\startdata
Open Cluster Common Name & \texttt{Common\_Name} & - & The common name of the open cluster to which the star belongs \\
Open Cluster Designation & \texttt{Name} & - & Cluster designation used to query the \citet{HuntReffert23} catalog \\
Open Cluster Age & \texttt{Age} & Myr & Ages come from previous literature (see Table \ref{tab:clusters}). \\
Gaia DR3 Source ID & \texttt{GaiaDR3} & - & - \\
TESS Input Catalog ID & \texttt{TICID} & - & - \\
Right Ascension (R.A.) & \texttt{ra} & $^\circ$ & - \\
Declination (Dec) & \texttt{dec} & $^\circ$ & - \\
Apparent Gaia $G$ magnitude & \texttt{Gmag} & mag & - \\
Apparent TESS magnitude & \texttt{Tmag} & mag & - \\
Absolute Gaia $G$ magnitude & \texttt{abs\_G} & mag & - \\
$V$-band cluster extinction & \texttt{Av} & mag & Median $A_V$ from \citet{HuntReffert23} \\
Gaia $G_{\rm BP}-G_{\rm RP}$ color & \texttt{bp\_rp} & mag & - \\
Dereddened Gaia $G_{\rm BP}-G_{\rm RP}$ color & \texttt{bp\_rp\_0} & mag & - \\
Gaia RUWE & \texttt{ruwe} & - & - \\
$v\sin i$ & \texttt{vsini} & ${\rm km\,s^{-1}}$ & Projected Rotational Velocity from Gaia \texttt{vbroad} or \citet{2023ApJ...946L..10B} \\
$v\sin i$ Error & \texttt{vsini\_err} & ${\rm km\,s^{-1}}$ & Error on the projected rotational velocity \\
Effective Temperature & \texttt{Teff} & K & From the TESS Input Catalog \citep{2019AJ....158..138S}\\
Effective Temperature Error & \texttt{s\_Teff} & K & ''\\
log Luminosity & \texttt{Lum} & ${\rm L_\odot}$ & ''\\
log Luminosity Error & \texttt{s\_Lum} & ${\rm L_\odot}$ & ''\\
TESS Sectors & \texttt{num\_sectors} & - & Number of TESS sectors used \\
log Skew & \texttt{skew} & - & log skewness of the distribution of peak amplitudes (\S\ref{subsec:identification}) \\
Maximum Amplitude & \texttt{max\_amp} & ppt & - \\
Maximum SNR & \texttt{max\_snr} & - & SNR of the maximum amplitude \\
$\delta$ Scuti & \texttt{dsct} & - & Boolean flag, 1 for $\delta$ Scuti stars, 0 for non-detections \\
\bottomrule
\enddata
\end{deluxetable*}

Our result in Figure \ref{fig:occur_v_teff_v_age} suggests that hotter $\delta$ Scuti stars may stop pulsating earlier than cooler stars. In hotter pulsators, the convective helium ionization zone is thinner and located closer to the surface, potentially allowing helium to deplete more rapidly than in cooler stars \citep{2005A&A...443..627T}. Part of the apparent decline at higher temperatures may also reflect basic stellar evolution, since A-type stars cool as they age and ultimately evolve off the main sequence.

Our results in Figures \ref{fig:vbroad_hists} and \ref{fig:frac_v_vbroad} demonstrate that age and rotation are closely linked in shaping pulsator occurrence, supporting the idea that helium settling can suppress $\delta$ Scuti pulsations. Helium diffusion into deeper stellar layers can be counteracted by rotational mixing, which helps helium remain in the ionization zone over time \citep{2004A&A...425..591H}. In this scenario, rapid rotators may continue pulsating longer than slow rotators. This is precisely the trend observed in Figure \ref{fig:vbroad_hists}. In older clusters, where pulsator occurrence is lower, the remaining $\delta$ Scuti stars rotate more rapidly than pulsators in younger clusters, on average. Figure \ref{fig:frac_v_vbroad} also shows that more slowly rotating pulsators tend to be younger than more rapidly rotating pulsators. While helium diffusion may be the dominant mechanism underlying our results, other excitation mechanisms, such as turbulent pressure, can also contribute to $\delta$ Scuti pulsations \citep{2014ApJ...796..118A}.

\section{Conclusions \& Outlook}
We have measured the $\delta$ Scuti pulsator occurrence in 20 open clusters to study how the occurrence varies with age, rotation, $T_{\rm eff}$, and metallicity. We obtained the following results:

\begin{enumerate}
    \item We found that the pulsator occurrence decreases with age, which suggests that $\delta$ Scuti stars can stop pulsating over time. The pulsator occurrence in open clusters younger than 200 Myr is consistently high, which an average occurrence of 88$\pm3$\%. The pulsator occurrence in clusters older than 200 Myr is lower, with an average occurrence of $62\pm3$\%.
    \item A dependence between the $\delta$ Scuti pulsator occurrence and metallicity may exist, with more metal-rich stars maintaining their $\delta$ Scuti pulsations over a longer period of time than more metal-poor stars. This trend is tentative, as we did not have many metal-rich or metal-poor clusters in our sample.
    \item We found a possible $T_{\rm eff}$-age dependence on the $\delta$ Scuti occurrence, where hotter $\delta$ Scuti stars ($\gtrsim$ 8500 K) may stop pulsating earlier than their cooler counterparts.
    \item We found a link between age, rotation, and the $\delta$ Scuti pulsator occurrence, where $\delta$ Scuti stars in older open clusters with lower pulsator occurrences are more rapidly rotating than pulsators in younger open clusters with higher pulsator occurrences. This shows that rapid rotation is crucial in maintaining $\delta$ Scuti pulsations over the main sequence lifetimes of these stars.
\end{enumerate}

Much more work can be done with $\delta$ Scuti stars in open clusters. With TESS, we limited this study to open clusters within 500 pc. Many more populous open clusters exist at further distances \citep{HuntReffert23,HuntReffert24}, which may be excellent candidates for testing and extending the results shown in this work. The $\delta$ Scuti pulsator occurrence may be measured in more distant open clusters using the upcoming PLAnetary Transists and Oscillations of Stars high-precision space photometer \citep[PLATO;][]{rauer2024platomission}, many of which exist within the first field to be observed by PLATO for two years \citep[LOPS2;][]{2025A&A...694A.313N}. Furthermore, we plan to model the effects of helium diffusion on the $\delta$ Scuti pulsator occurrence in the future, using the Yale Rotating Evolution Code
\citep[YREC;][]{1964ApJ...140..524L,1989ApJ...338..424P, 2008Ap&SS.316...31D, 2013ApJ...776...67V, pinsonneault2026yrecstellarevolutioncode}.

Beyond pulsator occurrence, $\delta$ Scuti stars with regular mode spacings enable precise age estimates for young open clusters ($\lesssim 200$ Myr) through asteroseismic modeling \citep[e.g.][]{2020Natur.581..147B, 2023MNRAS.526.3779M, 2026MNRAS.545f2001G}. Age-dating $\delta$ Scuti stars older than 1 Gyr with 1D stellar models has proven to be difficult \citep{2025A&A...703A..31P}. How other asteroseismic properties, such as pulsation amplitude and frequency, evolve with age and rotation remains to be explored using open clusters.

\section*{Acknowledgments}
I.B. and D.H. acknowledge support from the National Aeronautics and Space Administration (80NSSC22K0781). Y.L. acknowledges support from the National Aeronautics and Space Administration (80NSSC25K7904) and the Beatrice Watson Parrent Fellowship. T.R.B and S.J.M. were supported by the Australian Research Council through Laureate Fellowship FL220100117 and Future Fellowship FT2100100485.

This paper includes data collected with the TESS mission, obtained from the MAST data archive at the Space Telescope Science Institute (STScI). Funding for the TESS mission is provided by the NASA Explorer Program. STScI is operated by the Association of Universities for Research in Astronomy, Inc., under NASA contract NAS 5–26555.

\section*{Data Availability Statement}
The data presented in this paper are available via a machine-readable table. This work used data from the TIC catalog \citep{TIC_catalog}, FFI photometry from TESS Sectors 27 and above \citep{TESS_Team2022-av}, as well as TESS Long \citep[2-minute;][]{TESS_Team2021-gx} and Fast \citep[20-second;][]{TESS_Team2021-gk} cadence targeted photometry. The TGLC light curves used for NGC 3532 are available at DOI: \dataset[10.5281/zenodo.17402160]{https://doi.org/10.5281/zenodo.17402160} 


\bibliographystyle{aasjournal}
\bibliography{export-bibtex.bib}

@ARTICLE{2015JATIS...1a4003R,
       author = {{Ricker}, George R. and {Winn}, Joshua N. and {Vanderspek}, Roland and {Latham}, David W. and {Bakos}, G{\'a}sp{\'a}r {\'A}. and {Bean}, Jacob L. and {Berta-Thompson}, Zachory K. and {Brown}, Timothy M. and {Buchhave}, Lars and {Butler}, Nathaniel R. and {Butler}, R. Paul and {Chaplin}, William J. and {Charbonneau}, David and {Christensen-Dalsgaard}, J{\o}rgen and {Clampin}, Mark and {Deming}, Drake and {Doty}, John and {De Lee}, Nathan and {Dressing}, Courtney and {Dunham}, Edward W. and {Endl}, Michael and {Fressin}, Francois and {Ge}, Jian and {Henning}, Thomas and {Holman}, Matthew J. and {Howard}, Andrew W. and {Ida}, Shigeru and {Jenkins}, Jon M. and {Jernigan}, Garrett and {Johnson}, John Asher and {Kaltenegger}, Lisa and {Kawai}, Nobuyuki and {Kjeldsen}, Hans and {Laughlin}, Gregory and {Levine}, Alan M. and {Lin}, Douglas and {Lissauer}, Jack J. and {MacQueen}, Phillip and {Marcy}, Geoffrey and {McCullough}, Peter R. and {Morton}, Timothy D. and {Narita}, Norio and {Paegert}, Martin and {Palle}, Enric and {Pepe}, Francesco and {Pepper}, Joshua and {Quirrenbach}, Andreas and {Rinehart}, Stephen A. and {Sasselov}, Dimitar and {Sato}, Bun'ei and {Seager}, Sara and {Sozzetti}, Alessandro and {Stassun}, Keivan G. and {Sullivan}, Peter and {Szentgyorgyi}, Andrew and {Torres}, Guillermo and {Udry}, Stephane and {Villasenor}, Joel},
        title = "{Transiting Exoplanet Survey Satellite (TESS)}",
      journal = {Journal of Astronomical Telescopes, Instruments, and Systems},
         year = 2015,
        month = jan,
       volume = {1},
          eid = {014003},
        pages = {014003},
          doi = {10.1117/1.JATIS.1.1.014003},
       adsurl = {https://ui.adsabs.harvard.edu/abs/2015JATIS...1a4003R}
}

@MISC{2018ascl.soft12013L,
   author = {{Lightkurve Collaboration} and {Cardoso}, J.~V.~d.~M. and
             {Hedges}, C. and {Gully-Santiago}, M. and {Saunders}, N. and
             {Cody}, A.~M. and {Barclay}, T. and {Hall}, O. and
             {Sagear}, S. and {Turtelboom}, E. and {Zhang}, J. and
             {Tzanidakis}, A. and {Mighell}, K. and {Coughlin}, J. and
             {Bell}, K. and {Berta-Thompson}, Z. and {Williams}, P. and
             {Dotson}, J. and {Barentsen}, G.},
    title = "{Lightkurve: Kepler and TESS time series analysis in Python}",
howpublished = {Astrophysics Source Code Library},
     year = 2018,
    month = dec,
archivePrefix = "ascl",
   eprint = {1812.013},
   adsurl = {http://adsabs.harvard.edu/abs/2018ascl.soft12013L},
}

@article{mcinnes2017hdbscan,
  title={hdbscan: Hierarchical density based clustering},
  author={McInnes, Leland and Healy, John and Astels, Steve},
  journal={The Journal of Open Source Software},
  volume={2},
  number={11},
  pages={205},
  year={2017}
}

@ARTICLE{2013ApJ...776...67V,
       author = {{van Saders}, Jennifer L. and {Pinsonneault}, Marc H.},
        title = "{Fast Star, Slow Star; Old Star, Young Star: Subgiant Rotation as a Population and Stellar Physics Diagnostic}",
      journal = {\apj},
         year = 2013,
        month = oct,
       volume = {776},
       number = {2},
          eid = {67},
        pages = {67},
          doi = {10.1088/0004-637X/776/2/67},
archivePrefix = {arXiv},
       eprint = {1306.3701},
 primaryClass = {astro-ph.SR},
       adsurl = {https://ui.adsabs.harvard.edu/abs/2013ApJ...776...67V}
}

@ARTICLE{2004A&A...414L..17D,
       author = {{Dupret}, M. -A. and {Grigahc{\`e}ne}, A. and {Garrido}, R. and {Gabriel}, M. and {Scuflaire}, R.},
        title = "{Theoretical instability strips for {\ensuremath{\delta}} Scuti and {\ensuremath{\gamma}} Doradus stars}",
      journal = {\aap},
         year = 2004,
        month = jan,
       volume = {414},
        pages = {L17-L20},
          doi = {10.1051/0004-6361:20031740},
       adsurl = {https://ui.adsabs.harvard.edu/abs/2004A&A...414L..17D}
}

@ARTICLE{2014MNRAS.444..102K,
       author = {{Kurtz}, Donald W. and {Saio}, Hideyuki and {Takata}, Masao and {Shibahashi}, Hiromoto and {Murphy}, Simon J. and {Sekii}, Takashi},
        title = "{Asteroseismic measurement of surface-to-core rotation in a main-sequence A star, KIC 11145123}",
      journal = {\mnras},
         year = 2014,
        month = oct,
       volume = {444},
       number = {1},
        pages = {102-116},
          doi = {10.1093/mnras/stu1329},
archivePrefix = {arXiv},
       eprint = {1405.0155},
 primaryClass = {astro-ph.SR},
       adsurl = {https://ui.adsabs.harvard.edu/abs/2014MNRAS.444..102K}
}

@ARTICLE{2020Natur.581..147B,
       author = {{Bedding}, Timothy R. and {Murphy}, Simon J. and {Hey}, Daniel R. and {Huber}, Daniel and {Li}, Tanda and {Smalley}, Barry and {Stello}, Dennis and {White}, Timothy R. and {Ball}, Warrick H. and {Chaplin}, William J. and {Colman}, Isabel L. and {Fuller}, Jim and {Gaidos}, Eric and {Harbeck}, Daniel R. and {Hermes}, J.~J. and {Holdsworth}, Daniel L. and {Li}, Gang and {Li}, Yaguang and {Mann}, Andrew W. and {Reese}, Daniel R. and {Sekaran}, Sanjay and {Yu}, Jie and {Antoci}, Victoria and {Bergmann}, Christoph and {Brown}, Timothy M. and {Howard}, Andrew W. and {Ireland}, Michael J. and {Isaacson}, Howard and {Jenkins}, Jon M. and {Kjeldsen}, Hans and {McCully}, Curtis and {Rabus}, Markus and {Rains}, Adam D. and {Ricker}, George R. and {Tinney}, Christopher G. and {Vanderspek}, Roland K.},
        title = "{Very regular high-frequency pulsation modes in young intermediate-mass stars}",
      journal = {\nat},
         year = 2020,
        month = may,
       volume = {581},
       number = {7807},
        pages = {147-151},
          doi = {10.1038/s41586-020-2226-8},
archivePrefix = {arXiv},
       eprint = {2005.06157},
 primaryClass = {astro-ph.SR},
       adsurl = {https://ui.adsabs.harvard.edu/abs/2020Natur.581..147B}
}

@ARTICLE{2023MNRAS.526.3779M,
       author = {{Murphy}, Simon J. and {Bedding}, Timothy R. and {Gautam}, Anuj and {Joyce}, Meridith},
        title = "{A grid of 200 000 models of young {\ensuremath{\delta}} Scuti stars using MESA and GYRE}",
      journal = {\mnras},
         year = 2023,
        month = dec,
       volume = {526},
       number = {3},
        pages = {3779-3795},
          doi = {10.1093/mnras/stad2849},
archivePrefix = {arXiv},
       eprint = {2306.13142},
 primaryClass = {astro-ph.SR},
       adsurl = {https://ui.adsabs.harvard.edu/abs/2023MNRAS.526.3779M}
}

@ARTICLE{2024ApJ...972..137G,
       author = {{Gootkin}, Keyan and {Hon}, Marc and {Huber}, Daniel and {Hey}, Daniel R. and {Bedding}, Timothy R. and {Murphy}, Simon J.},
        title = "{A New Catalog of 100,000 Variable TESS A-F Stars Reveals a Correlation between {\ensuremath{\delta}} Scuti Pulsator Fraction and Stellar Rotation}",
      journal = {\apj},
         year = 2024,
        month = sep,
       volume = {972},
       number = {2},
          eid = {137},
        pages = {137},
          doi = {10.3847/1538-4357/ad5282},
archivePrefix = {arXiv},
       eprint = {2405.19388},
 primaryClass = {astro-ph.SR},
       adsurl = {https://ui.adsabs.harvard.edu/abs/2024ApJ...972..137G}
}

@ARTICLE{2024A&A...690A.104D,
       author = {{D{\"u}rfeldt-Pedros}, O. and {Antoci}, V. and {Smalley}, B. and {Murphy}, S. and {Posilek}, N. and {Niemczura}, E.},
        title = "{Variability and stellar pulsation incidence in Am and Fm stars using TESS and Gaia data}",
      journal = {\aap},
         year = 2024,
        month = oct,
       volume = {690},
          eid = {A104},
        pages = {A104},
          doi = {10.1051/0004-6361/202349076},
archivePrefix = {arXiv},
       eprint = {2408.11657},
 primaryClass = {astro-ph.SR},
       adsurl = {https://ui.adsabs.harvard.edu/abs/2024A&A...690A.104D}
}

@ARTICLE{2024MNRAS.528.2464R,
       author = {{Read}, Amelie K. and {Bedding}, Timothy R. and {Mani}, Prasad and {Montet}, Benjamin T. and {Crawford}, Courtney and {Hey}, Daniel R. and {Li}, Yaguang and {Murphy}, Simon J. and {Pedersen}, May Gade and {Kruger}, Joachim},
        title = "{Identifying 850 {\ensuremath{\delta}} Scuti pulsators in a narrow Gaia colour range with TESS 10-min full-frame images}",
      journal = {\mnras},
         year = 2024,
        month = feb,
       volume = {528},
       number = {2},
        pages = {2464-2473},
          doi = {10.1093/mnras/stae165},
archivePrefix = {arXiv},
       eprint = {2401.07413},
 primaryClass = {astro-ph.SR},
       adsurl = {https://ui.adsabs.harvard.edu/abs/2024MNRAS.528.2464R}
}

@ARTICLE{2023ApJ...946L..10B,
       author = {{Bedding}, Timothy R. and {Murphy}, Simon J. and {Crawford}, Courtney and {Hey}, Daniel R. and {Huber}, Daniel and {Kjeldsen}, Hans and {Li}, Yaguang and {Mann}, Andrew W. and {Torres}, Guillermo and {White}, Timothy R. and {Zhou}, George},
        title = "{TESS Observations of the Pleiades Cluster: A Nursery for {\ensuremath{\delta}} Scuti Stars}",
      journal = {\apjl},
         year = 2023,
        month = mar,
       volume = {946},
       number = {1},
          eid = {L10},
        pages = {L10},
          doi = {10.3847/2041-8213/acc17a},
archivePrefix = {arXiv},
       eprint = {2212.12087},
 primaryClass = {astro-ph.SR},
       adsurl = {https://ui.adsabs.harvard.edu/abs/2023ApJ...946L..10B}
}

@ARTICLE{2022MNRAS.513..374P,
       author = {{Pamos Ortega}, David and {Garc{\'\i}a Hern{\'a}ndez}, Antonio and {Su{\'a}rez}, Juan Carlos and {Pascual Granado}, Javier and {Barcel{\'o} Forteza}, Sebasti{\`a} and {Rod{\'o}n}, Jos{\'e} Ram{\'o}n},
        title = "{Determining the seismic age of the young open cluster {\ensuremath{\alpha}} Per using {\ensuremath{\delta}} Scuti stars}",
      journal = {\mnras},
         year = 2022,
        month = jun,
       volume = {513},
       number = {1},
        pages = {374-388},
          doi = {10.1093/mnras/stac864},
archivePrefix = {arXiv},
       eprint = {2203.14256},
 primaryClass = {astro-ph.SR},
       adsurl = {https://ui.adsabs.harvard.edu/abs/2022MNRAS.513..374P}
}

@ARTICLE{2024A&A...686A.142L,
       author = {{Li}, Gang and {Aerts}, Conny and {Bedding}, Timothy R. and {Fritzewski}, Dario J. and {Murphy}, Simon J. and {Van Reeth}, Timothy and {Montet}, Benjamin T. and {Jian}, Mingjie and {Mombarg}, Joey S.~G. and {Gossage}, Seth and {Sreenivas}, Kalarickal R.},
        title = "{Asteroseismology of the young open cluster NGC 2516. I. Photometric and spectroscopic observations}",
      journal = {\aap},
         year = 2024,
        month = jun,
       volume = {686},
          eid = {A142},
        pages = {A142},
          doi = {10.1051/0004-6361/202348901},
archivePrefix = {arXiv},
       eprint = {2311.16991},
 primaryClass = {astro-ph.SR},
       adsurl = {https://ui.adsabs.harvard.edu/abs/2024A&A...686A.142L}
}

@article{2023Gaia,
   title={GaiaData Release 3: Summary of the content and survey properties},
   volume={674},
   ISSN={1432-0746},
   url={http://dx.doi.org/10.1051/0004-6361/202243940},
   DOI={10.1051/0004-6361/202243940},
   journal={\aap},
   publisher={EDP Sciences},
   author={Vallenari, A. and Brown, A. G. A. and Prusti, T. and de Bruijne, J. H. J. and Arenou, F. and Babusiaux, C. and Biermann, M. and Creevey, O. L. and Ducourant, C. and Evans, D. W. and Eyer, L. and Guerra, R. and Hutton, A. and Jordi, C. and Klioner, S. A. and Lammers, U. L. and Lindegren, L. and Luri, X. and Mignard, F. and Panem, C. and Pourbaix, D. and Randich, S. and Sartoretti, P. and Soubiran, C. and Tanga, P. and Walton, N. A. and Bailer-Jones, C. A. L. and Bastian, U. and Drimmel, R. and Jansen, F. and Katz, D. and Lattanzi, M. G. and van Leeuwen, F. and Bakker, J. and Cacciari, C. and Castañeda, J. and De Angeli, F. and Fabricius, C. and Fouesneau, M. and Frémat, Y. and Galluccio, L. and Guerrier, A. and Heiter, U. and Masana, E. and Messineo, R. and Mowlavi, N. and Nicolas, C. and Nienartowicz, K. and Pailler, F. and Panuzzo, P. and Riclet, F. and Roux, W. and Seabroke, G. M. and Sordo, R. and Thévenin, F. and Gracia-Abril, G. and Portell, J. and Teyssier, D. and Altmann, M. and Andrae, R. and Audard, M. and Bellas-Velidis, I. and Benson, K. and Berthier, J. and Blomme, R. and Burgess, P. W. and Busonero, D. and Busso, G. and Cánovas, H. and Carry, B. and Cellino, A. and Cheek, N. and Clementini, G. and Damerdji, Y. and Davidson, M. and de Teodoro, P. and Nuñez Campos, M. and Delchambre, L. and Dell’Oro, A. and Esquej, P. and Fernández-Hernández, J. and Fraile, E. and Garabato, D. and García-Lario, P. and Gosset, E. and Haigron, R. and Halbwachs, J.-L. and Hambly, N. C. and Harrison, D. L. and Hernández, J. and Hestroffer, D. and Hodgkin, S. T. and Holl, B. and Janßen, K. and Jevardat de Fombelle, G. and Jordan, S. and Krone-Martins, A. and Lanzafame, A. C. and Löffler, W. and Marchal, O. and Marrese, P. M. and Moitinho, A. and Muinonen, K. and Osborne, P. and Pancino, E. and Pauwels, T. and Recio-Blanco, A. and Reylé, C. and Riello, M. and Rimoldini, L. and Roegiers, T. and Rybizki, J. and Sarro, L. M. and Siopis, C. and Smith, M. and Sozzetti, A. and Utrilla, E. and van Leeuwen, M. and Abbas, U. and Ábrahám, P. and Abreu Aramburu, A. and Aerts, C. and Aguado, J. J. and Ajaj, M. and Aldea-Montero, F. and Altavilla, G. and Álvarez, M. A. and Alves, J. and Anders, F. and Anderson, R. I. and Anglada Varela, E. and Antoja, T. and Baines, D. and Baker, S. G. and Balaguer-Núñez, L. and Balbinot, E. and Balog, Z. and Barache, C. and Barbato, D. and Barros, M. and Barstow, M. A. and Bartolomé, S. and Bassilana, J.-L. and Bauchet, N. and Becciani, U. and Bellazzini, M. and Berihuete, A. and Bernet, M. and Bertone, S. and Bianchi, L. and Binnenfeld, A. and Blanco-Cuaresma, S. and Blazere, A. and Boch, T. and Bombrun, A. and Bossini, D. and Bouquillon, S. and Bragaglia, A. and Bramante, L. and Breedt, E. and Bressan, A. and Brouillet, N. and Brugaletta, E. and Bucciarelli, B. and Burlacu, A. and Butkevich, A. G. and Buzzi, R. and Caffau, E. and Cancelliere, R. and Cantat-Gaudin, T. and Carballo, R. and Carlucci, T. and Carnerero, M. I. and Carrasco, J. M. and Casamiquela, L. and Castellani, M. and Castro-Ginard, A. and Chaoul, L. and Charlot, P. and Chemin, L. and Chiaramida, V. and Chiavassa, A. and Chornay, N. and Comoretto, G. and Contursi, G. and Cooper, W. J. and Cornez, T. and Cowell, S. and Crifo, F. and Cropper, M. and Crosta, M. and Crowley, C. and Dafonte, C. and Dapergolas, A. and David, M. and David, P. and de Laverny, P. and De Luise, F. and De March, R. and De Ridder, J. and de Souza, R. and de Torres, A. and del Peloso, E. F. and del Pozo, E. and Delbo, M. and Delgado, A. and Delisle, J.-B. and Demouchy, C. and Dharmawardena, T. E. and Di Matteo, P. and Diakite, S. and Diener, C. and Distefano, E. and Dolding, C. and Edvardsson, B. and Enke, H. and Fabre, C. and Fabrizio, M. and Faigler, S. and Fedorets, G. and Fernique, P. and Fienga, A. and Figueras, F. and Fournier, Y. and Fouron, C. and Fragkoudi, F. and Gai, M. and Garcia-Gutierrez, A. and Garcia-Reinaldos, M. and García-Torres, M. and Garofalo, A. and Gavel, A. and Gavras, P. and Gerlach, E. and Geyer, R. and Giacobbe, P. and Gilmore, G. and Girona, S. and Giuffrida, G. and Gomel, R. and Gomez, A. and González-Núñez, J. and González-Santamaría, I. and González-Vidal, J. J. and Granvik, M. and Guillout, P. and Guiraud, J. and Gutiérrez-Sánchez, R. and Guy, L. P. and Hatzidimitriou, D. and Hauser, M. and Haywood, M. and Helmer, A. and Helmi, A. and Sarmiento, M. H. and Hidalgo, S. L. and Hilger, T. and Hładczuk, N. and Hobbs, D. and Holland, G. and Huckle, H. E. and Jardine, K. and Jasniewicz, G. and Jean-Antoine Piccolo, A. and Jiménez-Arranz, Ó. and Jorissen, A. and Juaristi Campillo, J. and Julbe, F. and Karbevska, L. and Kervella, P. and Khanna, S. and Kontizas, M. and Kordopatis, G. and Korn, A. J. and Kóspál, Á and Kostrzewa-Rutkowska, Z. and Kruszyńska, K. and Kun, M. and Laizeau, P. and Lambert, S. and Lanza, A. F. and Lasne, Y. and Le Campion, J.-F. and Lebreton, Y. and Lebzelter, T. and Leccia, S. and Leclerc, N. and Lecoeur-Taibi, I. and Liao, S. and Licata, E. L. and Lindstrøm, H. E. P. and Lister, T. A. and Livanou, E. and Lobel, A. and Lorca, A. and Loup, C. and Madrero Pardo, P. and Magdaleno Romeo, A. and Managau, S. and Mann, R. G. and Manteiga, M. and Marchant, J. M. and Marconi, M. and Marcos, J. and Marcos Santos, M. M. S. and Marín Pina, D. and Marinoni, S. and Marocco, F. and Marshall, D. J. and Martin Polo, L. and Martín-Fleitas, J. M. and Marton, G. and Mary, N. and Masip, A. and Massari, D. and Mastrobuono-Battisti, A. and Mazeh, T. and McMillan, P. J. and Messina, S. and Michalik, D. and Millar, N. R. and Mints, A. and Molina, D. and Molinaro, R. and Molnár, L. and Monari, G. and Monguió, M. and Montegriffo, P. and Montero, A. and Mor, R. and Mora, A. and Morbidelli, R. and Morel, T. and Morris, D. and Muraveva, T. and Murphy, C. P. and Musella, I. and Nagy, Z. and Noval, L. and Ocaña, F. and Ogden, A. and Ordenovic, C. and Osinde, J. O. and Pagani, C. and Pagano, I. and Palaversa, L. and Palicio, P. A. and Pallas-Quintela, L. and Panahi, A. and Payne-Wardenaar, S. and Peñalosa Esteller, X. and Penttilä, A. and Pichon, B. and Piersimoni, A. M. and Pineau, F.-X. and Plachy, E. and Plum, G. and Poggio, E. and Prša, A. and Pulone, L. and Racero, E. and Ragaini, S. and Rainer, M. and Raiteri, C. M. and Rambaux, N. and Ramos, P. and Ramos-Lerate, M. and Re Fiorentin, P. and Regibo, S. and Richards, P. J. and Rios Diaz, C. and Ripepi, V. and Riva, A. and Rix, H.-W. and Rixon, G. and Robichon, N. and Robin, A. C. and Robin, C. and Roelens, M. and Rogues, H. R. O. and Rohrbasser, L. and Romero-Gómez, M. and Rowell, N. and Royer, F. and Ruz Mieres, D. and Rybicki, K. A. and Sadowski, G. and Sáez Núñez, A. and Sagristà Sellés, A. and Sahlmann, J. and Salguero, E. and Samaras, N. and Sanchez Gimenez, V. and Sanna, N. and Santoveña, R. and Sarasso, M. and Schultheis, M. and Sciacca, E. and Segol, M. and Segovia, J. C. and Ségransan, D. and Semeux, D. and Shahaf, S. and Siddiqui, H. I. and Siebert, A. and Siltala, L. and Silvelo, A. and Slezak, E. and Slezak, I. and Smart, R. L. and Snaith, O. N. and Solano, E. and Solitro, F. and Souami, D. and Souchay, J. and Spagna, A. and Spina, L. and Spoto, F. and Steele, I. A. and Steidelmüller, H. and Stephenson, C. A. and Süveges, M. and Surdej, J. and Szabados, L. and Szegedi-Elek, E. and Taris, F. and Taylor, M. B. and Teixeira, R. and Tolomei, L. and Tonello, N. and Torra, F. and Torra, J. and Torralba Elipe, G. and Trabucchi, M. and Tsounis, A. T. and Turon, C. and Ulla, A. and Unger, N. and Vaillant, M. V. and van Dillen, E. and van Reeven, W. and Vanel, O. and Vecchiato, A. and Viala, Y. and Vicente, D. and Voutsinas, S. and Weiler, M. and Wevers, T. and Wyrzykowski, Ł. and Yoldas, A. and Yvard, P. and Zhao, H. and Zorec, J. and Zucker, S. and Zwitter, T.},
   year={2023},
   month=jun, pages={A1} }

@ARTICLE{2020RNAAS...4..204H,
       author = {{Huang}, Chelsea X. and {Vanderburg}, Andrew and {P{\'a}l}, Andras and {Sha}, Lizhou and {Yu}, Liang and {Fong}, Willie and {Fausnaugh}, Michael and {Shporer}, Avi and {Guerrero}, Natalia and {Vanderspek}, Roland and {Ricker}, George},
        title = "{Photometry of 10 Million Stars from the First Two Years of TESS Full Frame Images: Part I}",
      journal = {Research Notes of the American Astronomical Society},
         year = 2020,
        month = nov,
       volume = {4},
       number = {11},
          eid = {204},
        pages = {204},
          doi = {10.3847/2515-5172/abca2e},
archivePrefix = {arXiv},
       eprint = {2011.06459},
 primaryClass = {astro-ph.EP},
       adsurl = {https://ui.adsabs.harvard.edu/abs/2020RNAAS...4..204H}
}

@ARTICLE{2020RNAAS...4..201C,
       author = {{Caldwell}, Douglas A. and {Tenenbaum}, Peter and {Twicken}, Joseph D. and {Jenkins}, Jon M. and {Ting}, Eric and {Smith}, Jeffrey C. and {Hedges}, Christina and {Fausnaugh}, Michael M. and {Rose}, Mark and {Burke}, Christopher},
        title = "{TESS Science Processing Operations Center FFI Target List Products}",
      journal = {Research Notes of the American Astronomical Society},
         year = 2020,
        month = nov,
       volume = {4},
       number = {11},
          eid = {201},
        pages = {201},
          doi = {10.3847/2515-5172/abc9b3},
archivePrefix = {arXiv},
       eprint = {2011.05495},
 primaryClass = {astro-ph.EP},
       adsurl = {https://ui.adsabs.harvard.edu/abs/2020RNAAS...4..201C}
}

@ARTICLE{2023AJ....165...71H,
       author = {{Han}, Te and {Brandt}, Timothy D.},
        title = "{TESS-Gaia Light Curve: A PSF-based TESS FFI Light-curve Product}",
      journal = {\aj},
         year = 2023,
        month = feb,
       volume = {165},
       number = {2},
          eid = {71},
        pages = {71},
          doi = {10.3847/1538-3881/acaaa7},
archivePrefix = {arXiv},
       eprint = {2301.03704},
 primaryClass = {astro-ph.IM},
       adsurl = {https://ui.adsabs.harvard.edu/abs/2023AJ....165...71H}
}

@ARTICLE{HuntReffert23,
       author = {{Hunt}, Emily L. and {Reffert}, Sabine},
        title = "{Improving the open cluster census. II. An all-sky cluster catalogue with Gaia DR3}",
      journal = {\aap},
         year = 2023,
        month = may,
       volume = {673},
          eid = {A114},
        pages = {A114},
          doi = {10.1051/0004-6361/202346285},
archivePrefix = {arXiv},
       eprint = {2303.13424},
 primaryClass = {astro-ph.GA},
       adsurl = {https://ui.adsabs.harvard.edu/abs/2023A&A...673A.114H}
}

@ARTICLE{HuntReffert24,
       author = {{Hunt}, Emily L. and {Reffert}, Sabine},
        title = "{Improving the open cluster census. III. Using cluster masses, radii, and dynamics to create a cleaned open cluster catalogue}",
      journal = {\aap},
         year = 2024,
        month = jun,
       volume = {686},
          eid = {A42},
        pages = {A42},
          doi = {10.1051/0004-6361/202348662},
archivePrefix = {arXiv},
       eprint = {2403.05143},
 primaryClass = {astro-ph.GA},
       adsurl = {https://ui.adsabs.harvard.edu/abs/2024A&A...686A..42H}
}

@ARTICLE{2019MNRAS.485.2380M,
       author = {{Murphy}, Simon J. and {Hey}, Daniel and {Van Reeth}, Timothy and {Bedding}, Timothy R.},
        title = "{Gaia-derived luminosities of Kepler A/F stars and the pulsator fraction across the {\ensuremath{\delta}} Scuti instability strip}",
      journal = {\mnras},
         year = 2019,
        month = may,
       volume = {485},
       number = {2},
        pages = {2380-2400},
          doi = {10.1093/mnras/stz590},
archivePrefix = {arXiv},
       eprint = {1903.00015},
 primaryClass = {astro-ph.SR},
       adsurl = {https://ui.adsabs.harvard.edu/abs/2019MNRAS.485.2380M}
}

@ARTICLE{2019AJ....158..138S,
       author = {{Stassun}, Keivan G. and {Oelkers}, Ryan J. and {Paegert}, Martin and {Torres}, Guillermo and {Pepper}, Joshua and {De Lee}, Nathan and {Collins}, Kevin and {Latham}, David W. and {Muirhead}, Philip S. and {Chittidi}, Jay and {Rojas-Ayala}, B{\'a}rbara and {Fleming}, Scott W. and {Rose}, Mark E. and {Tenenbaum}, Peter and {Ting}, Eric B. and {Kane}, Stephen R. and {Barclay}, Thomas and {Bean}, Jacob L. and {Brassuer}, C.~E. and {Charbonneau}, David and {Ge}, Jian and {Lissauer}, Jack J. and {Mann}, Andrew W. and {McLean}, Brian and {Mullally}, Susan and {Narita}, Norio and {Plavchan}, Peter and {Ricker}, George R. and {Sasselov}, Dimitar and {Seager}, S. and {Sharma}, Sanjib and {Shiao}, Bernie and {Sozzetti}, Alessandro and {Stello}, Dennis and {Vanderspek}, Roland and {Wallace}, Geoff and {Winn}, Joshua N.},
        title = "{The Revised TESS Input Catalog and Candidate Target List}",
      journal = {\aj},
         year = 2019,
        month = oct,
       volume = {158},
       number = {4},
          eid = {138},
        pages = {138},
          doi = {10.3847/1538-3881/ab3467},
archivePrefix = {arXiv},
       eprint = {1905.10694},
 primaryClass = {astro-ph.SR},
       adsurl = {https://ui.adsabs.harvard.edu/abs/2019AJ....158..138S}
}

@ARTICLE{2004A&A...425..591H,
       author = {{Huang}, R.~Q.},
        title = "{On rotational mixing in stars}",
      journal = {\aap},
         year = 2004,
        month = oct,
       volume = {425},
        pages = {591-594},
          doi = {10.1051/0004-6361:20034245},
       adsurl = {https://ui.adsabs.harvard.edu/abs/2004A&A...425..591H}
}

@ARTICLE{1970ApJ...162..597B,
       author = {{Breger}, Michel},
        title = "{Metallic-Line a Stars and Pulsation}",
      journal = {\apj},
         year = 1970,
        month = nov,
       volume = {162},
        pages = {597},
          doi = {10.1086/150691},
       adsurl = {https://ui.adsabs.harvard.edu/abs/1970ApJ...162..597B}
}

@ARTICLE{1974NInfo..32..104P,
       author = {{Pamjatnykh}, A.~A.},
        title = "{Diffusion of chemical elements in stellar envelopes}",
      journal = {Nauchnye Informatsii},
         year = 1974,
        month = jan,
       volume = {32},
        pages = {104},
       adsurl = {https://ui.adsabs.harvard.edu/abs/1974NInfo..32..104P}
}

@ARTICLE{2022MNRAS.516.2080B,
       author = {{Barac}, Natascha and {Bedding}, Timothy R. and {Murphy}, Simon J. and {Hey}, Daniel R.},
        title = "{Revisiting bright {\ensuremath{\delta}} Scuti stars and their period-luminosity relation with TESS and Gaia DR3}",
      journal = {\mnras},
         year = 2022,
        month = oct,
       volume = {516},
       number = {2},
        pages = {2080-2094},
          doi = {10.1093/mnras/stac2132},
archivePrefix = {arXiv},
       eprint = {2207.00343},
 primaryClass = {astro-ph.SR},
       adsurl = {https://ui.adsabs.harvard.edu/abs/2022MNRAS.516.2080B}
}

@ARTICLE{1973A&A....23..221B,
       author = {{Baglin}, A. and {Breger}, M. and {Chevalier}, C. and {Hauck}, B. and {Le Contel}, J.~M. and {Sareyan}, J.~P. and {Valtier}, J.~C.},
        title = "{Delta Scuti stars.}",
      journal = {\aap},
         year = 1973,
        month = mar,
       volume = {23},
        pages = {221},
       adsurl = {https://ui.adsabs.harvard.edu/abs/1973A&A....23..221B}
}

@ARTICLE{2005A&A...443..627T,
       author = {{Th{\'e}ado}, S. and {Vauclair}, S. and {Cunha}, M.~S.},
        title = "{Helium settling and mass loss in magnetic Ap stars. I. The chemical stratification}",
      journal = {\aap},
         year = 2005,
        month = nov,
       volume = {443},
       number = {2},
        pages = {627-641},
          doi = {10.1051/0004-6361:20052933},
       adsurl = {https://ui.adsabs.harvard.edu/abs/2005A&A...443..627T}
}

@ARTICLE{2016A&A...589A.140D,
       author = {{Deal}, Morgan and {Richard}, Olivier and {Vauclair}, Sylvie},
        title = "{Hydrodynamical instabilities induced by atomic diffusion in A stars and their consequences}",
      journal = {\aap},
         year = 2016,
        month = may,
       volume = {589},
          eid = {A140},
        pages = {A140},
          doi = {10.1051/0004-6361/201628180},
archivePrefix = {arXiv},
       eprint = {1604.01241},
 primaryClass = {astro-ph.SR},
       adsurl = {https://ui.adsabs.harvard.edu/abs/2016A&A...589A.140D}
}

@ARTICLE{2024MNRAS.534.3022M,
       author = {{Murphy}, Simon J. and {Bedding}, Timothy R. and {Gautam}, Anuj and {Kerr}, Ronan P. and {Mani}, Prasad},
        title = "{The {\ensuremath{\delta}} Scuti stars of the Cep-Her Complex - I. Pulsator fraction, rotation, asteroseismic large spacings, and the {\ensuremath{\nu}}$_{max}$ relation}",
      journal = {\mnras},
         year = 2024,
        month = nov,
       volume = {534},
       number = {4},
        pages = {3022-3039},
          doi = {10.1093/mnras/stae2226},
archivePrefix = {arXiv},
       eprint = {2409.13135},
 primaryClass = {astro-ph.SR},
       adsurl = {https://ui.adsabs.harvard.edu/abs/2024MNRAS.534.3022M}
}

@ARTICLE{2023A&A...675A.167P,
       author = {{Pamos Ortega}, D. and {Mirouh}, G.~M. and {Garc{\'\i}a Hern{\'a}ndez}, A. and {Su{\'a}rez Yanes}, J.~C. and {Barcel{\'o} Forteza}, S.},
        title = "{Dating young open clusters using {\ensuremath{\delta}} Scuti stars. Results for Trumpler 10 and Praesepe}",
      journal = {\aap},
         year = 2023,
        month = jul,
       volume = {675},
          eid = {A167},
        pages = {A167},
          doi = {10.1051/0004-6361/202346323},
archivePrefix = {arXiv},
       eprint = {2306.02791},
 primaryClass = {astro-ph.SR},
       adsurl = {https://ui.adsabs.harvard.edu/abs/2023A&A...675A.167P}
}

@article{Villanova_2009,
   title={A spectroscopic study of the open cluster NGC 6475 (M 7): Chemical abundances from stars in the rangeTeff = 4500–10 000 K},
   volume={504},
   ISSN={1432-0746},
   url={http://dx.doi.org/10.1051/0004-6361/200811507},
   DOI={10.1051/0004-6361/200811507},
   number={3},
   journal={\aap},
   publisher={EDP Sciences},
   author={Villanova, S. and Carraro, G. and Saviane, I.},
   year={2009},
   month=jul, pages={845–852} }

@ARTICLE{2010AN....331..989G,
       author = {{Grigahc{\`e}ne}, A. and {Uytterhoeven}, K. and {Antoci}, V. and {Balona}, L. and {Catanzaro}, G. and {Daszy{\'n}ska-Daszkiewicz}, J. and {Guzik}, J.~A. and {Handler}, G. and {Houdek}, G. and {Kurtz}, D.~W. and {Marconi}, M. and {Monteiro}, M.~J.~P.~F.~G. and {Moya}, A. and {Ripepi}, V. and {Su{\'a}rez}, J. -C. and {Borucki}, W.~J. and {Brown}, T.~M. and {Christensen-Dalsgaard}, J. and {Gilliland}, R.~L. and {Jenkins}, J.~M. and {Kjeldsen}, H. and {Koch}, D. and {Bernabei}, S. and {Bradley}, P. and {Breger}, M. and {Di Criscienzo}, M. and {Dupret}, M. -A. and {Garc{\'\i}a}, R.~A. and {Garc{\'\i}a Hern{\'a}ndez}, A. and {Jackiewicz}, J. and {Kaiser}, A. and {Lehmann}, H. and {Mart{\'\i}n-Ruiz}, S. and {Mathias}, P. and {Molenda-{\.Z}akowicz}, J. and {Nemec}, J.~M. and {Nuspl}, J. and {Papar{\'o}}, M. and {Roth}, M. and {Szab{\'o}}, R. and {Suran}, M.~D. and {Ventura}, R.},
        title = "{Kepler observations: Light shed on the hybrid {\ensuremath{\gamma}} Doradus - {\ensuremath{\delta}} Scuti pulsation phenomenon}",
      journal = {Astronomische Nachrichten},
         year = 2010,
        month = dec,
       volume = {331},
        pages = {989},
          doi = {10.1002/asna.201011443},
       adsurl = {https://ui.adsabs.harvard.edu/abs/2010AN....331..989G}
}

@ARTICLE{2018FrASS...5...43B,
       author = {{Balona}, Luis A.},
        title = "{Pulsation in Intermediate-Mass Stars}",
      journal = {Frontiers in Astronomy and Space Sciences},
         year = 2018,
        month = dec,
       volume = {5},
          eid = {43},
        pages = {43},
          doi = {10.3389/fspas.2018.00043},
       adsurl = {https://ui.adsabs.harvard.edu/abs/2018FrASS...5...43B}
}

@ARTICLE{2012A&A...537A.120Z,
       author = {{Zorec}, J. and {Royer}, F.},
        title = "{Rotational velocities of A-type stars. IV. Evolution of rotational velocities}",
      journal = {\aap},
         year = 2012,
        month = jan,
       volume = {537},
          eid = {A120},
        pages = {A120},
          doi = {10.1051/0004-6361/201117691},
archivePrefix = {arXiv},
       eprint = {1201.2052},
 primaryClass = {astro-ph.SR},
       adsurl = {https://ui.adsabs.harvard.edu/abs/2012A&A...537A.120Z}
}

@INCOLLECTION{1980LNP...125...22D,
       author = {{Dziembowski}, W.},
        title = "{Delta Scuti variables - The link between giant- and dwarf-type pulsators}",
    booktitle = {Nonradial and Nonlinear Stellar Pulsation},
         year = 1980,
       editor = {{Hill}, H.~A. and {Dziembowski}, W.~A.},
       volume = {125},
        pages = {22-33},
          doi = {10.1007/3-540-09994-8_2},
       adsurl = {https://ui.adsabs.harvard.edu/abs/1980LNP...125...22D}
}

@ARTICLE{2015A&A...579A.116O,
       author = {{Ouazzani}, R. -M. and {Roxburgh}, I.~W. and {Dupret}, M. -A.},
        title = "{Pulsations of rapidly rotating stars. II. Realistic modelling for intermediate-mass stars}",
      journal = {\aap},
         year = 2015,
        month = jul,
       volume = {579},
          eid = {A116},
        pages = {A116},
          doi = {10.1051/0004-6361/201525734},
archivePrefix = {arXiv},
       eprint = {1505.01088},
 primaryClass = {astro-ph.SR},
       adsurl = {https://ui.adsabs.harvard.edu/abs/2015A&A...579A.116O}
}

@ARTICLE{2021arXiv210709479G,
       author = {{Guzik}, Joyce Ann and {Jackiewicz}, Jason and {Catanzaro}, Giovanni and {Soukup}, Michael S.},
        title = "{Characterizing Variability in Bright Metallic-Line A (Am) Stars Using Data from the NASA TESS Spacecraft}",
      journal = {arXiv e-prints},
         year = 2021,
        month = jul,
          eid = {arXiv:2107.09479},
        pages = {arXiv:2107.09479},
          doi = {10.48550/arXiv.2107.09479},
archivePrefix = {arXiv},
       eprint = {2107.09479},
 primaryClass = {astro-ph.SR},
       adsurl = {https://ui.adsabs.harvard.edu/abs/2021arXiv210709479G}
}

@ARTICLE{2011AJ....141..115C,
       author = {{Clem}, James L. and {Landolt}, Arlo U. and {Hoard}, D.~W. and {Wachter}, Stefanie},
        title = "{Deep, Wide-field CCD Photometry for the Open Cluster NGC 3532}",
      journal = {\aj},
         year = 2011,
        month = apr,
       volume = {141},
       number = {4},
          eid = {115},
        pages = {115},
          doi = {10.1088/0004-6256/141/4/115},
archivePrefix = {arXiv},
       eprint = {1101.3268},
 primaryClass = {astro-ph.SR},
       adsurl = {https://ui.adsabs.harvard.edu/abs/2011AJ....141..115C}
}

@ARTICLE{1962AJ.....67..471K,
       author = {{King}, Ivan},
        title = "{The structure of star clusters. I. an empirical density law}",
      journal = {\aj},
         year = 1962,
        month = oct,
       volume = {67},
        pages = {471},
          doi = {10.1086/108756},
       adsurl = {https://ui.adsabs.harvard.edu/abs/1962AJ.....67..471K}
}

@article{astropy:2013,
Adsurl = {http://adsabs.harvard.edu/abs/2013A%26A...558A..33A},
Archiveprefix = {arXiv},
Author = {{Astropy Collaboration} and {Robitaille}, T.~P. and {Tollerud}, E.~J. and {Greenfield}, P. and {Droettboom}, M. and {Bray}, E. and {Aldcroft}, T. and {Davis}, M. and {Ginsburg}, A. and {Price-Whelan}, A.~M. and {Kerzendorf}, W.~E. and {Conley}, A. and {Crighton}, N. and {Barbary}, K. and {Muna}, D. and {Ferguson}, H. and {Grollier}, F. and {Parikh}, M.~M. and {Nair}, P.~H. and {Unther}, H.~M. and {Deil}, C. and {Woillez}, J. and {Conseil}, S. and {Kramer}, R. and {Turner}, J.~E.~H. and {Singer}, L. and {Fox}, R. and {Weaver}, B.~A. and {Zabalza}, V. and {Edwards}, Z.~I. and {Azalee Bostroem}, K. and {Burke}, D.~J. and {Casey}, A.~R. and {Crawford}, S.~M. and {Dencheva}, N. and {Ely}, J. and {Jenness}, T. and {Labrie}, K. and {Lim}, P.~L. and {Pierfederici}, F. and {Pontzen}, A. and {Ptak}, A. and {Refsdal}, B. and {Servillat}, M. and {Streicher}, O.},
Doi = {10.1051/0004-6361/201322068},
Eid = {A33},
Eprint = {1307.6212},
Journal = {\aap},
Month = oct,
Pages = {A33},
Primaryclass = {astro-ph.IM},
Title = {{Astropy: A community Python package for astronomy}},
Volume = 558,
Year = 2013}

@ARTICLE{astropy:2018,
       author = {{Astropy Collaboration} and {Price-Whelan}, A.~M. and
         {Sip{\H{o}}cz}, B.~M. and {G{\"u}nther}, H.~M. and {Lim}, P.~L. and
         {Crawford}, S.~M. and {Conseil}, S. and {Shupe}, D.~L. and
         {Craig}, M.~W. and {Dencheva}, N. and {Ginsburg}, A. and {Vand
        erPlas}, J.~T. and {Bradley}, L.~D. and {P{\'e}rez-Su{\'a}rez}, D. and
         {de Val-Borro}, M. and {Aldcroft}, T.~L. and {Cruz}, K.~L. and
         {Robitaille}, T.~P. and {Tollerud}, E.~J. and {Ardelean}, C. and
         {Babej}, T. and {Bach}, Y.~P. and {Bachetti}, M. and {Bakanov}, A.~V. and
         {Bamford}, S.~P. and {Barentsen}, G. and {Barmby}, P. and
         {Baumbach}, A. and {Berry}, K.~L. and {Biscani}, F. and {Boquien}, M. and
         {Bostroem}, K.~A. and {Bouma}, L.~G. and {Brammer}, G.~B. and
         {Bray}, E.~M. and {Breytenbach}, H. and {Buddelmeijer}, H. and
         {Burke}, D.~J. and {Calderone}, G. and {Cano Rodr{\'\i}guez}, J.~L. and
         {Cara}, M. and {Cardoso}, J.~V.~M. and {Cheedella}, S. and {Copin}, Y. and
         {Corrales}, L. and {Crichton}, D. and {D'Avella}, D. and {Deil}, C. and
         {Depagne}, {\'E}. and {Dietrich}, J.~P. and {Donath}, A. and
         {Droettboom}, M. and {Earl}, N. and {Erben}, T. and {Fabbro}, S. and
         {Ferreira}, L.~A. and {Finethy}, T. and {Fox}, R.~T. and
         {Garrison}, L.~H. and {Gibbons}, S.~L.~J. and {Goldstein}, D.~A. and
         {Gommers}, R. and {Greco}, J.~P. and {Greenfield}, P. and
         {Groener}, A.~M. and {Grollier}, F. and {Hagen}, A. and {Hirst}, P. and
         {Homeier}, D. and {Horton}, A.~J. and {Hosseinzadeh}, G. and {Hu}, L. and
         {Hunkeler}, J.~S. and {Ivezi{\'c}}, {\v{Z}}. and {Jain}, A. and
         {Jenness}, T. and {Kanarek}, G. and {Kendrew}, S. and {Kern}, N.~S. and
         {Kerzendorf}, W.~E. and {Khvalko}, A. and {King}, J. and {Kirkby}, D. and
         {Kulkarni}, A.~M. and {Kumar}, A. and {Lee}, A. and {Lenz}, D. and
         {Littlefair}, S.~P. and {Ma}, Z. and {Macleod}, D.~M. and
         {Mastropietro}, M. and {McCully}, C. and {Montagnac}, S. and
         {Morris}, B.~M. and {Mueller}, M. and {Mumford}, S.~J. and {Muna}, D. and
         {Murphy}, N.~A. and {Nelson}, S. and {Nguyen}, G.~H. and
         {Ninan}, J.~P. and {N{\"o}the}, M. and {Ogaz}, S. and {Oh}, S. and
         {Parejko}, J.~K. and {Parley}, N. and {Pascual}, S. and {Patil}, R. and
         {Patil}, A.~A. and {Plunkett}, A.~L. and {Prochaska}, J.~X. and
         {Rastogi}, T. and {Reddy Janga}, V. and {Sabater}, J. and
         {Sakurikar}, P. and {Seifert}, M. and {Sherbert}, L.~E. and
         {Sherwood-Taylor}, H. and {Shih}, A.~Y. and {Sick}, J. and
         {Silbiger}, M.~T. and {Singanamalla}, S. and {Singer}, L.~P. and
         {Sladen}, P.~H. and {Sooley}, K.~A. and {Sornarajah}, S. and
         {Streicher}, O. and {Teuben}, P. and {Thomas}, S.~W. and
         {Tremblay}, G.~R. and {Turner}, J.~E.~H. and {Terr{\'o}n}, V. and
         {van Kerkwijk}, M.~H. and {de la Vega}, A. and {Watkins}, L.~L. and
         {Weaver}, B.~A. and {Whitmore}, J.~B. and {Woillez}, J. and
         {Zabalza}, V. and {Astropy Contributors}},
        title = "{The Astropy Project: Building an Open-science Project and Status of the v2.0 Core Package}",
      journal = {\aj},
         year = 2018,
        month = sep,
       volume = {156},
       number = {3},
          eid = {123},
        pages = {123},
          doi = {10.3847/1538-3881/aabc4f},
archivePrefix = {arXiv},
       eprint = {1801.02634},
 primaryClass = {astro-ph.IM},
       adsurl = {https://ui.adsabs.harvard.edu/abs/2018AJ....156..123A}
}

@ARTICLE{astropy:2022,
       author = {{Astropy Collaboration} and {Price-Whelan}, Adrian M. and {Lim}, Pey Lian and {Earl}, Nicholas and {Starkman}, Nathaniel and {Bradley}, Larry and {Shupe}, David L. and {Patil}, Aarya A. and {Corrales}, Lia and {Brasseur}, C.~E. and {N{"o}the}, Maximilian and {Donath}, Axel and {Tollerud}, Erik and {Morris}, Brett M. and {Ginsburg}, Adam and {Vaher}, Eero and {Weaver}, Benjamin A. and {Tocknell}, James and {Jamieson}, William and {van Kerkwijk}, Marten H. and {Robitaille}, Thomas P. and {Merry}, Bruce and {Bachetti}, Matteo and {G{"u}nther}, H. Moritz and {Aldcroft}, Thomas L. and {Alvarado-Montes}, Jaime A. and {Archibald}, Anne M. and {B{'o}di}, Attila and {Bapat}, Shreyas and {Barentsen}, Geert and {Baz{'a}n}, Juanjo and {Biswas}, Manish and {Boquien}, M{'e}d{'e}ric and {Burke}, D.~J. and {Cara}, Daria and {Cara}, Mihai and {Conroy}, Kyle E. and {Conseil}, Simon and {Craig}, Matthew W. and {Cross}, Robert M. and {Cruz}, Kelle L. and {D'Eugenio}, Francesco and {Dencheva}, Nadia and {Devillepoix}, Hadrien A.~R. and {Dietrich}, J{"o}rg P. and {Eigenbrot}, Arthur Davis and {Erben}, Thomas and {Ferreira}, Leonardo and {Foreman-Mackey}, Daniel and {Fox}, Ryan and {Freij}, Nabil and {Garg}, Suyog and {Geda}, Robel and {Glattly}, Lauren and {Gondhalekar}, Yash and {Gordon}, Karl D. and {Grant}, David and {Greenfield}, Perry and {Groener}, Austen M. and {Guest}, Steve and {Gurovich}, Sebastian and {Handberg}, Rasmus and {Hart}, Akeem and {Hatfield-Dodds}, Zac and {Homeier}, Derek and {Hosseinzadeh}, Griffin and {Jenness}, Tim and {Jones}, Craig K. and {Joseph}, Prajwel and {Kalmbach}, J. Bryce and {Karamehmetoglu}, Emir and {Ka{l}uszy{'n}ski}, Miko{l}aj and {Kelley}, Michael S.~P. and {Kern}, Nicholas and {Kerzendorf}, Wolfgang E. and {Koch}, Eric W. and {Kulumani}, Shankar and {Lee}, Antony and {Ly}, Chun and {Ma}, Zhiyuan and {MacBride}, Conor and {Maljaars}, Jakob M. and {Muna}, Demitri and {Murphy}, N.~A. and {Norman}, Henrik and {O'Steen}, Richard and {Oman}, Kyle A. and {Pacifici}, Camilla and {Pascual}, Sergio and {Pascual-Granado}, J. and {Patil}, Rohit R. and {Perren}, Gabriel I. and {Pickering}, Timothy E. and {Rastogi}, Tanuj and {Roulston}, Benjamin R. and {Ryan}, Daniel F. and {Rykoff}, Eli S. and {Sabater}, Jose and {Sakurikar}, Parikshit and {Salgado}, Jes{'u}s and {Sanghi}, Aniket and {Saunders}, Nicholas and {Savchenko}, Volodymyr and {Schwardt}, Ludwig and {Seifert-Eckert}, Michael and {Shih}, Albert Y. and {Jain}, Anany Shrey and {Shukla}, Gyanendra and {Sick}, Jonathan and {Simpson}, Chris and {Singanamalla}, Sudheesh and {Singer}, Leo P. and {Singhal}, Jaladh and {Sinha}, Manodeep and {Sip{H{o}}cz}, Brigitta M. and {Spitler}, Lee R. and {Stansby}, David and {Streicher}, Ole and {{{S}}umak}, Jani and {Swinbank}, John D. and {Taranu}, Dan S. and {Tewary}, Nikita and {Tremblay}, Grant R. and {Val-Borro}, Miguel de and {Van Kooten}, Samuel J. and {Vasovi{'c}}, Zlatan and {Verma}, Shresth and {de Miranda Cardoso}, Jos{'e} Vin{'i}cius and {Williams}, Peter K.~G. and {Wilson}, Tom J. and {Winkel}, Benjamin and {Wood-Vasey}, W.~M. and {Xue}, Rui and {Yoachim}, Peter and {Zhang}, Chen and {Zonca}, Andrea and {Astropy Project Contributors}},
        title = "{The Astropy Project: Sustaining and Growing a Community-oriented Open-source Project and the Latest Major Release (v5.0) of the Core Package}",
      journal = {\apj},
         year = 2022,
        month = aug,
       volume = {935},
       number = {2},
          eid = {167},
        pages = {167},
          doi = {10.3847/1538-4357/ac7c74},
archivePrefix = {arXiv},
       eprint = {2206.14220},
 primaryClass = {astro-ph.IM},
       adsurl = {https://ui.adsabs.harvard.edu/abs/2022ApJ...935..167A}
}

@ARTICLE{2016ApJS..222....8D,
       author = {{Dotter}, Aaron},
        title = "{MESA Isochrones and Stellar Tracks (MIST) 0: Methods for the Construction of Stellar Isochrones}",
      journal = {\apjs},
         year = 2016,
        month = jan,
       volume = {222},
       number = {1},
          eid = {8},
        pages = {8},
          doi = {10.3847/0067-0049/222/1/8},
archivePrefix = {arXiv},
       eprint = {1601.05144},
 primaryClass = {astro-ph.SR},
       adsurl = {https://ui.adsabs.harvard.edu/abs/2016ApJS..222....8D}
}

@ARTICLE{2016ApJ...823..102C,
       author = {{Choi}, Jieun and {Dotter}, Aaron and {Conroy}, Charlie and {Cantiello}, Matteo and {Paxton}, Bill and {Johnson}, Benjamin D.},
        title = "{Mesa Isochrones and Stellar Tracks (MIST). I. Solar-scaled Models}",
      journal = {\apj},
         year = 2016,
        month = jun,
       volume = {823},
       number = {2},
          eid = {102},
        pages = {102},
          doi = {10.3847/0004-637X/823/2/102},
archivePrefix = {arXiv},
       eprint = {1604.08592},
 primaryClass = {astro-ph.SR},
       adsurl = {https://ui.adsabs.harvard.edu/abs/2016ApJ...823..102C}
}

@ARTICLE{2023A&A...674A...8F,
       author = {{Fr{\'e}mat}, Y. and {Royer}, F. and {Marchal}, O. and {Blomme}, R. and {Sartoretti}, P. and {Guerrier}, A. and {Panuzzo}, P. and {Katz}, D. and {Seabroke}, G.~M. and {Th{\'e}venin}, F. and {Cropper}, M. and {Benson}, K. and {Damerdji}, Y. and {Haigron}, R. and {Lobel}, A. and {Smith}, M. and {Baker}, S.~G. and {Chemin}, L. and {David}, M. and {Dolding}, C. and {Gosset}, E. and {Jan{\ss}en}, K. and {Jasniewicz}, G. and {Plum}, G. and {Samaras}, N. and {Snaith}, O. and {Soubiran}, C. and {Vanel}, O. and {Zorec}, J. and {Zwitter}, T. and {Brouillet}, N. and {Caffau}, E. and {Crifo}, F. and {Fabre}, C. and {Fragkoudi}, F. and {Huckle}, H.~E. and {Lasne}, Y. and {Leclerc}, N. and {Mastrobuono-Battisti}, A. and {Jean-Antoine Piccolo}, A. and {Viala}, Y.},
        title = "{Gaia Data Release 3. Properties of the line-broadening parameter derived with the Radial Velocity Spectrometer (RVS)}",
      journal = {\aap},
         year = 2023,
        month = jun,
       volume = {674},
          eid = {A8},
        pages = {A8},
          doi = {10.1051/0004-6361/202243809},
archivePrefix = {arXiv},
       eprint = {2206.10986},
 primaryClass = {astro-ph.SR},
       adsurl = {https://ui.adsabs.harvard.edu/abs/2023A&A...674A...8F}
}

@ARTICLE{2018A&A...616A...5C,
       author = {{Cropper}, M. and {Katz}, D. and {Sartoretti}, P. and {Prusti}, T. and {de Bruijne}, J.~H.~J. and {Chassat}, F. and {Charvet}, P. and {Boyadjian}, J. and {Perryman}, M. and {Sarri}, G. and {Gare}, P. and {Erdmann}, M. and {Munari}, U. and {Zwitter}, T. and {Wilkinson}, M. and {Arenou}, F. and {Vallenari}, A. and {G{\'o}mez}, A. and {Panuzzo}, P. and {Seabroke}, G. and {Allende Prieto}, C. and {Benson}, K. and {Marchal}, O. and {Huckle}, H. and {Smith}, M. and {Dolding}, C. and {Jan{\ss}en}, K. and {Viala}, Y. and {Blomme}, R. and {Baker}, S. and {Boudreault}, S. and {Crifo}, F. and {Soubiran}, C. and {Fr{\'e}mat}, Y. and {Jasniewicz}, G. and {Guerrier}, A. and {Guy}, L.~P. and {Turon}, C. and {Jean-Antoine-Piccolo}, A. and {Th{\'e}venin}, F. and {David}, M. and {Gosset}, E. and {Damerdji}, Y.},
        title = "{Gaia Data Release 2. Gaia Radial Velocity Spectrometer}",
      journal = {\aap},
         year = 2018,
        month = aug,
       volume = {616},
          eid = {A5},
        pages = {A5},
          doi = {10.1051/0004-6361/201832763},
archivePrefix = {arXiv},
       eprint = {1804.09369},
 primaryClass = {astro-ph.IM},
       adsurl = {https://ui.adsabs.harvard.edu/abs/2018A&A...616A...5C}
}

@article{Kaye_1999,
   title={γ Doradus Stars: Defining a New Class of Pulsating Variables},
   volume={111},
   ISSN={1538-3873},
   url={http://dx.doi.org/10.1086/316399},
   DOI={10.1086/316399},
   number={761},
   journal={Publications of the Astronomical Society of the Pacific},
   publisher={IOP Publishing},
   author={Kaye, Anthony B. and Handler, Gerald and Krisciunas, Kevin and Poretti, Ennio and Zerbi, Filippo M.},
   year={1999},
   month=jul, pages={840–844} }

@misc{liu2025revisitingopenclusters200,
      title={Revisiting open clusters within 200 pc in the solar neighbourhood with Gaia DR3}, 
      author={Penghui Liu and Min Fang and Yue-Lin Sming Tsai and Xiaoying Pang and Fan Wang and Xiaoting Fu},
      year={2025},
      eprint={2504.08179},
      archivePrefix={arXiv},
      primaryClass={astro-ph.SR},
      url={https://arxiv.org/abs/2504.08179}, 
}

@ARTICLE{2010arXiv1007.3176H,
       author = {{Hareter}, M. and {Reegen}, P. and {Miglio}, A. and {Montalban}, J. and {Kaiser}, A. and {Dekany}, I. and {Guenther}, E. and {Poretti}, E. and {Mathias}, P. and {Weiss}, W.},
        title = "{Gamma Dor and Gamma Dor - Delta Sct Hybrid Stars In The CoRoT LRa01}",
      journal = {arXiv e-prints},
         year = 2010,
        month = jul,
          eid = {arXiv:1007.3176},
        pages = {arXiv:1007.3176},
          doi = {10.48550/arXiv.1007.3176},
archivePrefix = {arXiv},
       eprint = {1007.3176},
 primaryClass = {astro-ph.SR},
       adsurl = {https://ui.adsabs.harvard.edu/abs/2010arXiv1007.3176H}
}

@ARTICLE{2011A&A...534A.125U,
       author = {{Uytterhoeven}, K. and {Moya}, A. and {Grigahc{\`e}ne}, A. and {Guzik}, J.~A. and {Guti{\'e}rrez-Soto}, J. and {Smalley}, B. and {Handler}, G. and {Balona}, L.~A. and {Niemczura}, E. and {Fox Machado}, L. and {Benatti}, S. and {Chapellier}, E. and {Tkachenko}, A. and {Szab{\'o}}, R. and {Su{\'a}rez}, J.~C. and {Ripepi}, V. and {Pascual}, J. and {Mathias}, P. and {Mart{\'\i}n-Ru{\'\i}z}, S. and {Lehmann}, H. and {Jackiewicz}, J. and {Hekker}, S. and {Gruberbauer}, M. and {Garc{\'\i}a}, R.~A. and {Dumusque}, X. and {D{\'\i}az-Fraile}, D. and {Bradley}, P. and {Antoci}, V. and {Roth}, M. and {Leroy}, B. and {Murphy}, S.~J. and {De Cat}, P. and {Cuypers}, J. and {Kjeldsen}, H. and {Christensen-Dalsgaard}, J. and {Breger}, M. and {Pigulski}, A. and {Kiss}, L.~L. and {Still}, M. and {Thompson}, S.~E. and {van Cleve}, J.},
        title = "{The Kepler characterization of the variability among A- and F-type stars. I. General overview}",
      journal = {\aap},
         year = 2011,
        month = oct,
       volume = {534},
          eid = {A125},
        pages = {A125},
          doi = {10.1051/0004-6361/201117368},
archivePrefix = {arXiv},
       eprint = {1107.0335},
 primaryClass = {astro-ph.SR},
       adsurl = {https://ui.adsabs.harvard.edu/abs/2011A&A...534A.125U}
}

@ARTICLE{2005A&A...435..927D,
       author = {{Dupret}, M. -A. and {Grigahc{\`e}ne}, A. and {Garrido}, R. and {Gabriel}, M. and {Scuflaire}, R.},
        title = "{Convection-pulsation coupling. II. Excitation and stabilization mechanisms in {\ensuremath{\delta}} Sct and {\ensuremath{\gamma}} Dor stars}",
      journal = {\aap},
         year = 2005,
        month = jun,
       volume = {435},
       number = {3},
        pages = {927-939},
          doi = {10.1051/0004-6361:20041817},
       adsurl = {https://ui.adsabs.harvard.edu/abs/2005A&A...435..927D}
}

@ARTICLE{2023A&A...674A..36G,
       author = {{Gaia Collaboration} and {De Ridder}, J. and {Ripepi}, V. and {Aerts}, C. and {Palaversa}, L. and {Eyer}, L. and {Holl}, B. and {Audard}, M. and {Rimoldini}, L. and {Brown}, A.~G.~A. and {Vallenari}, A. and {Prusti}, T. and {de Bruijne}, J.~H.~J. and {Arenou}, F. and {Babusiaux}, C. and {Biermann}, M. and {Creevey}, O.~L. and {Ducourant}, C. and {Evans}, D.~W. and {Guerra}, R. and {Hutton}, A. and {Jordi}, C. and {Klioner}, S.~A. and {Lammers}, U.~L. and {Lindegren}, L. and {Luri}, X. and {Mignard}, F. and {Panem}, C. and {Pourbaix}, D. and {Randich}, S. and {Sartoretti}, P. and {Soubiran}, C. and {Tanga}, P. and {Walton}, N.~A. and {Bailer-Jones}, C.~A.~L. and {Bastian}, U. and {Drimmel}, R. and {Jansen}, F. and {Katz}, D. and {Lattanzi}, M.~G. and {van Leeuwen}, F. and {Bakker}, J. and {Cacciari}, C. and {Casta{\~n}eda}, J. and {De Angeli}, F. and {Fabricius}, C. and {Fouesneau}, M. and {Fr{\'e}mat}, Y. and {Galluccio}, L. and {Guerrier}, A. and {Heiter}, U. and {Masana}, E. and {Messineo}, R. and {Mowlavi}, N. and {Nicolas}, C. and {Nienartowicz}, K. and {Pailler}, F. and {Panuzzo}, P. and {Riclet}, F. and {Roux}, W. and {Seabroke}, G.~M. and {Sordo}, R. and {Th{\'e}venin}, F. and {Gracia-Abril}, G. and {Portell}, J. and {Teyssier}, D. and {Altmann}, M. and {Andrae}, R. and {Bellas-Velidis}, I. and {Benson}, K. and {Berthier}, J. and {Blomme}, R. and {Burgess}, P.~W. and {Busonero}, D. and {Busso}, G. and {C{\'a}novas}, H. and {Carry}, B. and {Cellino}, A. and {Cheek}, N. and {Clementini}, G. and {Damerdji}, Y. and {Davidson}, M. and {de Teodoro}, P. and {Nu{\~n}ez Campos}, M. and {Delchambre}, L. and {Dell'Oro}, A. and {Esquej}, P. and {Fern{\'a}ndez-Hern{\'a}ndez}, J. and {Fraile}, E. and {Garabato}, D. and {Garc{\'\i}a-Lario}, P. and {Gosset}, E. and {Haigron}, R. and {Halbwachs}, J. -L. and {Hambly}, N.~C. and {Harrison}, D.~L. and {Hern{\'a}ndez}, J. and {Hestroffer}, D. and {Hilger}, T. and {Hodgkin}, S.~T. and {Jan{\ss}en}, K. and {Jevardat de Fombelle}, G. and {Jordan}, S. and {Krone-Martins}, A. and {Lanzafame}, A.~C. and {L{\"o}ffler}, W. and {Marchal}, O. and {Marrese}, P.~M. and {Moitinho}, A. and {Muinonen}, K. and {Osborne}, P. and {Pancino}, E. and {Pauwels}, T. and {Recio-Blanco}, A. and {Reyl{\'e}}, C. and {Riello}, M. and {Roegiers}, T. and {Rybizki}, J. and {Sarro}, L.~M. and {Siopis}, C. and {Smith}, M. and {Sozzetti}, A. and {Utrilla}, E. and {van Leeuwen}, M. and {Abbas}, U. and {{\'A}brah{\'a}m}, P. and {Abreu Aramburu}, A. and {Aguado}, J.~J. and {Ajaj}, M. and {Aldea-Montero}, F. and {Altavilla}, G. and {{\'A}lvarez}, M.~A. and {Alves}, J. and {Anders}, F. and {Anderson}, R.~I. and {Anglada Varela}, E. and {Antoja}, T. and {Baines}, D. and {Baker}, S.~G. and {Balaguer-N{\'u}{\~n}ez}, L. and {Balbinot}, E. and {Balog}, Z. and {Barache}, C. and {Barbato}, D. and {Barros}, M. and {Barstow}, M.~A. and {Bartolom{\'e}}, S. and {Bassilana}, J. -L. and {Bauchet}, N. and {Becciani}, U. and {Bellazzini}, M. and {Berihuete}, A. and {Bernet}, M. and {Bertone}, S. and {Bianchi}, L. and {Binnenfeld}, A. and {Blanco-Cuaresma}, S. and {Boch}, T. and {Bombrun}, A. and {Bossini}, D. and {Bouquillon}, S. and {Bragaglia}, A. and {Bramante}, L. and {Breedt}, E. and {Bressan}, A. and {Brouillet}, N. and {Brugaletta}, E. and {Bucciarelli}, B. and {Burlacu}, A. and {Butkevich}, A.~G. and {Buzzi}, R. and {Caffau}, E. and {Cancelliere}, R. and {Cantat-Gaudin}, T. and {Carballo}, R. and {Carlucci}, T. and {Carnerero}, M.~I. and {Carrasco}, J.~M. and {Casamiquela}, L. and {Castellani}, M. and {Castro-Ginard}, A. and {Chaoul}, L. and {Charlot}, P. and {Chemin}, L. and {Chiaramida}, V. and {Chiavassa}, A. and {Chornay}, N. and {Comoretto}, G. and {Contursi}, G. and {Cooper}, W.~J. and {Cornez}, T. and {Cowell}, S. and {Crifo}, F. and {Cropper}, M. and {Crosta}, M. and {Crowley}, C. and {Dafonte}, C. and {Dapergolas}, A. and {David}, P. and {de Laverny}, P.},
        title = "{Gaia Data Release 3. Pulsations in main sequence OBAF-type stars}",
      journal = {\aap},
         year = 2023,
        month = jun,
       volume = {674},
          eid = {A36},
        pages = {A36},
          doi = {10.1051/0004-6361/202243767},
archivePrefix = {arXiv},
       eprint = {2206.06075},
 primaryClass = {astro-ph.SR},
       adsurl = {https://ui.adsabs.harvard.edu/abs/2023A&A...674A..36G}
}

@ARTICLE{2019ApJ...877..116W,
       author = {{Wang}, Shu and {Chen}, Xiaodian},
        title = "{The Optical to Mid-infrared Extinction Law Based on the APOGEE, Gaia DR2, Pan-STARRS1, SDSS, APASS, 2MASS, and WISE Surveys}",
      journal = {\apj},
         year = 2019,
        month = jun,
       volume = {877},
       number = {2},
          eid = {116},
        pages = {116},
          doi = {10.3847/1538-4357/ab1c61},
archivePrefix = {arXiv},
       eprint = {1904.04575},
 primaryClass = {astro-ph.GA},
       adsurl = {https://ui.adsabs.harvard.edu/abs/2019ApJ...877..116W}
}

@ARTICLE{2019A&A...622A.110F,
       author = {{Fritzewski}, D.~J. and {Barnes}, S.~A. and {James}, D.~J. and {Geller}, A.~M. and {Meibom}, S. and {Strassmeier}, K.~G.},
        title = "{Spectroscopic membership for the populous 300 Myr-old open cluster NGC 3532}",
      journal = {\aap},
         year = 2019,
        month = feb,
       volume = {622},
          eid = {A110},
        pages = {A110},
          doi = {10.1051/0004-6361/201833587},
archivePrefix = {arXiv},
       eprint = {1901.04507},
 primaryClass = {astro-ph.SR},
       adsurl = {https://ui.adsabs.harvard.edu/abs/2019A&A...622A.110F}
}

@ARTICLE{2018AJ....156..102S,
       author = {{Stassun}, Keivan G. and {Oelkers}, Ryan J. and {Pepper}, Joshua and {Paegert}, Martin and {De Lee}, Nathan and {Torres}, Guillermo and {Latham}, David W. and {Charpinet}, St{\'e}phane and {Dressing}, Courtney D. and {Huber}, Daniel and {Kane}, Stephen R. and {L{\'e}pine}, S{\'e}bastien and {Mann}, Andrew and {Muirhead}, Philip S. and {Rojas-Ayala}, B{\'a}rbara and {Silvotti}, Roberto and {Fleming}, Scott W. and {Levine}, Al and {Plavchan}, Peter},
        title = "{The TESS Input Catalog and Candidate Target List}",
      journal = {\aj},
         year = 2018,
        month = sep,
       volume = {156},
       number = {3},
          eid = {102},
        pages = {102},
          doi = {10.3847/1538-3881/aad050},
archivePrefix = {arXiv},
       eprint = {1706.00495},
 primaryClass = {astro-ph.EP},
       adsurl = {https://ui.adsabs.harvard.edu/abs/2018AJ....156..102S}
}
\clearpage

\end{document}